\documentclass[]{elsarticle}
\usepackage[latin9]{inputenc}
\usepackage{geometry}
\usepackage{boxhandler}
\usepackage{graphicx}
\usepackage{dcolumn}
\usepackage{bm}
\usepackage{epsfig}
\usepackage{color}
\usepackage{longtable}
\usepackage{cancel}
\usepackage{float}
\usepackage{bm}
\usepackage[french,english]{babel}
\usepackage[OT1,T1]{fontenc}
\usepackage[normalem]{ulem}
\usepackage{lscape}
\usepackage{multirow}

\usepackage{longtable}
\begin{document}
		\title{$\gamma$-ray emission in proton-induced nuclear reactions on $^{nat}$C
			and Mylar targets over the incident energy range, E$_{p}$ = 30 \textendash{}
			200 MeV. Astrophysical implications}
		\author [1] {Y. Rahma}
		\ead{youssoufrahma@yahoo.com}
		\author [1]{S. Ouichaoui\corref{cor1}}
		\cortext[cor1]{Corresponding author}
		\ead{souichaoui@usthb.dz}
		\ead{souichaoui@gmail.com}
		\author [2,3]{J. Kiener}
		\author[4,5]{E. A. Lawrie}
		\author[4,5]{J. J. Lawrie}
		\author[2,3]{V. Tatischeff}
		\author[1]{A. Belhout}
		\author[1]{D. Moussa}
		\author[1]{W. Yahia-Cherif}
		\author[6]{H. Benhabiles-Mezhoud}
		\author[4,7]{T. D. Bucher}
		\author[8]{T. R. S. Dinoko}
		\author[1]{A. Chafa}
		\author[4]{J. L. Conradie}
		\author[9]{S. Damache}
		\author[1]{M. Debabi}
		\author[2,3]{I. Deloncle}
		\author[4,5]{J. L. Easton}
		\author[10]{M. Fouka}
		\author[2,3]{C. Hamadache}
		\author[11,3]{F. Hammache}
		\author[4]{P. Jones}
		\author[4,7]{B. V. Kheswa}
		\author[5]{N. A. Khumalo}
		\author[5]{T. Lamula}
		\author[4,12,13]{S. N. T. Majola}
		\author[4,7]{J. Ndayishimye}
		\author[4,14]{D. Negi}
		\author[4,5]{S. P. Noncolela}
		\author[1]{S. Ouziane}
		\author[4,7]{P. Papka}
		\author[12]{S. Peterson}
		\author[15]{M. Kumar Raju}
		\author[16]{V. Ramanathan}
		\author[5]{B. M. Rebeiro}
		\author[11,3]{N. de S{\'e}r{\'e}ville}
		\author[5]{J. F. Sharpey-Schafer}
		\author[4,7]{O. Shirinda}
		\author[4,17]{M. Wiedeking}
		\author[7]{S. Wyngaardt}
		\address [1] {University of Sciences and Technology Houari Boumediene
			(USTHB), Laboratory of Nuclear Sciences and Radiation-Matter Interactions
			(SNIRM-DGRSDT), Faculty of Physics, P.O. Box 32, EL Alia, 16111 Bab
			Ezzouar, Algiers, Algeria}
		\address [2] {Centre de Sciences Nucl{\'e}aires et de Sciences de la
			Mati{\`e}re (CSNSM), CNRS-IN2P3 et Universit{\'e} de Paris-Sud, 91405 Orsay
			Campus, France}
		\address [3]{Present address: Universit\'e Paris-Saclay, CNRS/IN2P3, IJCLab, 91405 Orsay, France}
		\address [4]{iThemba LABS, National Research Foundation, P.O. Box 722, Somerset West 7129, South Africa}
		\address [5]{Department of Physics, University of the Western Cape, Private Bag X17, Bellville 7535, South Africa}
		\address [6]{Universit{\'e} M\textquoteright Hamed Bougara, Institut
			de G{\'e}nie Electrique et Electronique, 35 000 Boumerd{\'e}s, Algeria}
		\address [7]{Department of Physics, Stellenbosch University, Private
			Bag X1, Matieland 7602, South Africa}
		\address [8]{Department of Physical and Electrical Metrology, NMISA, Private Bag X34, Lynnwood Ridge, Pretoria 0040, South Africa}
		\address [9]{CRNA, 02 Boulevard Frantz Fanon, B.P. 399 Alger-gare, Algiers, Algeria}
		\address [10]{Center for Research in Astronomy, Astrophysics and Geophysics,
			B.P. 63, Algiers observatory, Bouzaeah, Algiers, Algeria}
		\address [11]{Institut de Physique Nucl{\'e}aire (IPN), CNRS-IN2P3 et
		Universit{\'e} Paris-Sud, 91405 Orsay Campus, France}
		\address [12]{Department of Physics, University of Cape Town, Private Bag X3, 7701 Rondebosch, South Africa}
		\address [13]{Department of Physics, University of Johannesburg, P.O. Box 524, Auckland Park 2006, South Africa}
		\address [14]{Department
			of Nuclear and Atomic Physics, Tata Institute of Fundamental Research, Mumbai
			400005, India}
		\address [15]{Department of Physics, GITAM School of Science, Vishakhapatnam-530045, India}
		\address [16]{Department of Radiography and Radiotherapy, Faculty of Allied Health Sciences, General Sir John Kotelawala Defence University, 
			Ratmalana, Sri Lanka}
		\address [17]{School of Physics, University of the Witwatersrand, Johannesburg 2050, South Africa}
			\begin{abstract}
				We have measured the $\gamma$-ray line production cross sections
				in proton-induced nuclear reactions on various target nuclei ($^{12}$C,
				$^{16}$O, $^{24}$Mg, $^{28}$Si, $^{56}$Fe) of chemical elements
				abundant in astrophysical sites (solar flares, the interstellar medium,
				cosmic compact objects) over the incident energy range of E$_{p}$
				= 30 \textendash{} 200 MeV. We carried out experimental campaigns
				in joint collaboration at the K = 200 separated sector cyclotron
				of iThemba LABS using a high-energy resolution, high-efficiency detection
				array composed of 8 Compton-suppressed clover detectors comprising
				32 HP-Ge crystals for recording the $\gamma$-ray energy spectra. In the
				current paper, we focus on $\gamma$-ray de-excitation lines produced
				in proton irradiations of $^{nat}$C and Mylar targets, in particular,
				on the prominent 4.439 and 6.129 MeV lines of $^{12}$C and $^{16}$O
				which are among the strongest lines emitted in solar flares and in
				interactions of low-energy cosmic rays (LECRs) with the gas and dust
				of the inner galaxy. We report new $\gamma$-ray production experimental
				cross section data for ten nuclear $\gamma$-ray lines that we compare
				to previous low-energy data sets from the literature, to the predictions
				of the TALYS code of nuclear reactions and to a semi-empirical
				compilation. In the first approach, performing calculations with default
				input parameters of TALYS we observed substantial deviations between
				the predicted cross sections and experimental data. Then, using modified
				optical model potential (OMP) and nuclear level deformation parameters
				as input data we generated theoretical excitation functions for the
				above two main lines fully consistent with experimental data. In contrast,
				the experimental data sets for the other eight analyzed lines from
				the two proton-irradiated targets exhibit significant deviations with
				the predicted cross section values. We also report line-shape experimental
				data for the line complex observed at E$_{\gamma}$ = 4.44 MeV in
				irradiations of the two targets. Finally, we emphasize the astrophysical
				implications of our results.
			\end{abstract}
			\begin{keyword}
				Nuclear reactions, Gamma-ray spectrometry, Gamma-ray line shape analysis, Gamma-ray production cross section, Solar flares, Interstellar medium.
			\end{keyword}
			\maketitle
			
			\section{Introduction}
			
			The last few decades have seen the achievement of considerable progress
			in the elucidation of the non-thermal acceleration of charged particles
			in magnetized astrophysical sites, their interactions and transport,
			as well as the induced radiation processes that provide crucial information
			on the properties of the energetic particles themselves and of the
			accelerating media \cite{ramaty1994gamma,murphy2007using,LorenzoSironiandAnatolySpitkovsky,benhabiles2013erratum,T.VieuS.GabiciV.Tatischeff}.
			In this context, the emission of nuclear $\gamma$-ray lines produced
			in violent collisions between highly accelerated ions and abundant
			nuclei in solar flares (SFs) and the interstellar medium (ISM) has
			been extensively studied via several complementary methods \cite{ramaty1979nuclear,kozlovsky2002nuclear,R.J.MurphyB.KozlovskyJ.Kiener2009,H.Benhabiles-Mezhoud2011,W.Yahia-Cherif2020}
			(and Refs therein). These studies included laboratory measurements
			of $\gamma$-ray cross sections, their comparisons to the predictions
			of nuclear reaction codes like EMPIRE2 \cite{M.HermanEMPIRE} and
			TALYS \cite{TALYS}, the modeling of $\gamma$-ray fluxes from astrophysical
			sites \cite{tatischeff2003EAS,V.Tatischeff2004,murphy2016evidence,tusnski2019self}
			and their comparison to observational data reported by space telescopes
			such as SMM \cite{murphy1991solar,share1998accelerated}, CGRO \cite{share1998accelerated},
			RHESSI \cite{lin2003rhessi}, INTEGRAL \cite{kiener2006properties,harris2007high}
			or FERMI \cite{Fermi_flares}. This allowed an increasing understanding
			with time of the nuclear collision processes at work in SFs and the
			ISM.
			
			In both astrophysical sites, the most important nuclear reactions are those induced by energetic protons and $\alpha$-particles
			on the nuclei of abundant chemical elements (H, He, C, N, O, Ne, Mg,
			Al, Si, S, Ca and Fe), and the inverse reactions between accelerated
			ions of the latter species and ambient hydrogen and helium \cite{ramaty1979nuclear,kozlovsky2002nuclear,R.J.MurphyB.KozlovskyJ.Kiener2009,H.Benhabiles-Mezhoud2011,W.Yahia-Cherif2020}.
			$\gamma$-ray lines are from various processes like the de-excitation
			of directly excited nuclei via inelastic scattering, the products
			of nuclear reactions (transfer, fusion-evaporation or spallation reactions)
			and synthesized radioactive nuclei. These reactions give rise, respectively,
			to a narrow and a broad component of strong lines in the observed
			complex $\gamma$-ray emission spectra from the SFs and ISM, in addition
			to a non-resolved component of weak lines from highly excited states
			forming a quasi-continuum \cite{Vestrand1999}. Consequently, the
			knowledge of $\gamma$-ray production cross sections over wide particle
			energy regimes, as present in solar flares and cosmic rays, extending
			from reaction thresholds up to hundreds of MeV/u is required.
			
			In a pioneer reference work, Ramaty et al. \cite{ramaty1979nuclear}
			established in 1979 a database of $\gamma$-ray line production cross
			sections reporting $\gamma$-ray experimental cross section data for
			main de-excitation lines from low-lying excited states produced in
			nuclear reactions induced by light ions (protons, $^{3}$He$^{+}$
			ions and $\alpha$-particles) on target nuclei from $^{4}$He to $^{56}$Fe
			abundant in astrophysical sites. Successively updated in 2002 by Kozlovsky
			et al.\textbf{ }\cite{kozlovsky2002nuclear} and in 2009 by Murphy
			et al. \cite{R.J.MurphyB.KozlovskyJ.Kiener2009}, this compilation
			provides extrapolations of existing cross section data to higher projectile
			energies not covered by experiment. It also predicts the variation
			trends of $\gamma$-ray excitation functions for numerous weak lines
			emitted by high-lying excited states, exclusively based on TALYS code
			\cite{TALYS} calculations.
			
			Laboratory measurements of these observables have been carried out
			mainly at the Washington \cite{Dyer81,Narayanaswamy81,Seamster84,Dyer85}
			and Orsay \cite{H.Benhabiles-Mezhoud2011,Belhout2007,Belhout2009,Benhabiles-Mezhoud2010,Kiener2008,Kiener1998}
			tandem accelerators for reactions induced on various target nuclei
			by protons and $\alpha$-particles with incident energies covering
			the ranges reaching up to E$_{p}$ = 26 and E$_{\alpha}$ = 40 MeV,
			respectively. Experimental data sets for some $\gamma$-ray lines
			have also been measured at cyclotron facilities for incident energies
			extending up to E$_{p}$ = 50 MeV and E$_{\alpha}$ = 40 MeV \cite{Lesko1988}
			(and references therein). More recently, new data sets for the strongest
			$\gamma$-ray lines produced in $\alpha$-particle-induced reactions
			on C, Mg, Si and Fe targets, extending the available data over the
			energy range, E$_{\alpha}$ = 50 - 90 MeV, have been measured by Kiener
			et al. \cite{Kiener2021PhyRevC} at the center for proton therapy
			of the Helmholtz-Zentrum in Berlin.
			
			The 4439 and 6129 keV transitions from the J$^{\pi}$ = 2$^{+}$ first
			excited state of $^{12}$C and the 3$^{-}$ second excited state of
			$^{16}$O to the 0$^{+}$ ground states of these nuclei are among
			the most intense lines in the $\gamma$-ray energy spectra from proton-induced
			reactions \cite{R.J.MurphyB.KozlovskyJ.Kiener2009}. Other intense
			lines from these reactions \cite{R.J.MurphyB.KozlovskyJ.Kiener2009}
			are the 511 keV line from the positron-electron annihilation and the
			847 keV, 1369 keV, 1634 and 1779 keV lines of $^{56}$Fe, $^{24}$Mg,
			$^{20}$Ne and $^{28}$Si, respectively, and the 2223-keV line from
			neutron capture on hydrogen, which is a very prominent line in solar
			flares.
			
			The 4439 and 6129 keV lines are produced in astrophysical sites mainly
			via inelastic proton and $\alpha$-particle scatterings off $^{12}$C
			and $^{16}$O nuclei, respectively, and (for the 4.439 MeV line) in
			the $^{16}$O(p, p$\alpha$)$^{12}$C and $^{16}$O($\alpha$, 2$\alpha$)$^{12}$C
			reactions and in interactions of accelerated $^{12}$C and $^{16}$O
			ions with abundant hydrogen and helium. Production cross sections
			for these two lines are available up to E$_{p}$ = 85 MeV \cite{Lang1987},
			and E$_{\alpha}$ = 90 MeV for the 4439 keV line of $^{12}$C \cite{Kiener2021PhyRevC}.
			However, experimental $\gamma$-ray production cross-section data
			yet remains lacking for higher proton energies. These two transitions
			are of great importance for line shape studies in particle-induced
			nuclear reactions, as pointed out by Kiener \cite{Kiener2019} who
			has recently performed a detailed and comprehensive line shape analysis
			on the 4439-keV $\gamma$-ray transition in $^{12}$C in terms of
			dominant nuclear reaction mechanisms. Indeed, the $\gamma$-ray line
			shape analysis could allow extracting important information on the
			properties of the accelerated particle populations, the underlying
			particle acceleration mechanism and the magneto-hydrodynamic structure
			of these sites \cite{Casse1995,Parizot1999}.
			
			We have measured $\gamma$-ray line production cross sections for
			30-200 MeV protons accelerated on various target nuclei of astrophysical
			concern at the separated sector cyclotron (SSC) of iThemba LABS (near
			Cape Town, in South Africa). We have reported part of the experimental
			data sets involving the Si, Mg and Fe targets in a recent publication
			{[}10{]}. This work extended previous experimental data for the main
			lines of these nuclei to higher proton energies up to E$_{p}$ = 66
			MeV and allowed us to obtain cross section data for some weaker lines.
			
			In the present paper, we report new experimental data for de-excitation
			$\gamma$-ray lines produced in proton-induced nuclear reactions on
			$^{nat}$C and Mylar targets over the incident energy range of E$_{p}$
			= 30 - 200 MeV. In section \textbf{\ref{sec:Experiments}} we provide
			experimental details on the set-up and procedure used. We present
			our data analysis and experimental results in section \textbf{\ref{sec:Data-analysis,-experimental}}.
			In section \textbf{\ref{sec:Comparison-of-our}} we compare the extracted
			$\gamma$-ray line production cross sections to previous data sets
			from the literature. Then, we compare them in section \textbf{\ref{sec:Comparison-of-experimental}}
			to values predicted by nuclear reaction models derived via TALYS code
			\cite{TALYS} calculations and by the compilation of Murphy, Kozlosky,
			Kiener and Share \cite{R.J.MurphyB.KozlovskyJ.Kiener2009}. Finally,
			a summary and perspectives are given in section \textbf{\ref{sec:Summary,-discussion-and}}
			where we also emphasize the astrophysical implications of our results.

			\section{Experiments\label{sec:Experiments}}
			
			One can find a detailed description of the set up and methods used
			in our recent paper \cite{W.Yahia-Cherif2020}.
			
			Since previous experimental $\gamma$-ray line production cross section
			data sets were available mostly for proton energies extending up to
			E$_{p}$ = 26 MeV \cite{H.Benhabiles-Mezhoud2011,Dyer81,Narayanaswamy81,
			Seamster84,Dyer85,Belhout2007,Belhout2009,Benhabiles-Mezhoud2010,
			Kiener2008,Kiener1998,Lesko1988},
			we decided to investigate the higher proton energy region of E$_{p}$
			= 30 \textendash{} 200 MeV, allowed by the SSC facility of iThemba
			LABS. We have thus undertaken experimental campaigns in joint scientific
			collaboration, taking data over the energy regions of E$_{p}$ = 30
			\textendash{} 66 MeV, 66 \textendash{} 125 MeV and 125 \textendash{}
			200 MeV in 2014 - 2016. More precisely we carried out measurements
			at proton beam energies of E$_{p}$ = 30, 42, 54 and 66 MeV in 2014,
			Ep = 66, 80, 95, 110 and 125 MeV in 2015, and E$_{p}$ = 66, 125, 150, 175,
			200 MeV in 2016. That is a total of 11 different beam energies, (the
			measurements at E$_{p}$ = 66 and 125 MeV being repeated for data
			checking). Proton beams were delivered by the cyclotron accelerator
			with current intensities in the range of I = 2 - 5 nA, and focused
			onto $^{nat}$C, Mylar (chemical composition C$_{10}$H$_{8}$O$_{4}$)
			and other (Mg, Al, Si, Ca and Fe) targets of interest. A well-manufactured,
			electrically isolated Faraday cup placed 3 m downstream of the reaction
			chamber (protected deep inside a thick concrete wall) served to stop
			the incident proton beam and to measure its integrated current with
			high precision of $\approx$ 1\%. The collected total beam charge
			was typically of $\approx$ 5 $\mu$C for a run of 1-hour duration.
			In all our experiments, we used self-supporting, natural or isotopically
			enriched solid targets that were prepared at iThemba LABS (Cape Town)
			and we brought some of them from the CSNSM-Orsay. Table \ref{tab:1}
			reports the properties of the $^{nat}$C and Mylar targets that had
			thicknesses of 8.4 and 7 mg cm$^{-2}$, respectively, while the other
			elemental targets had thicknesses in the range of 6 - 9.8 mg cm$^{-2}$
			(see Ref. \cite{W.Yahia-Cherif2020}). The targets were mounted on
			rectangular Al frames, then onto a common holder. An Al frame without
			target and a fluorescent Al$_{2}$O$_{3}$ beam
			viewer with a central 3 mm diameter hole devoted, respectively, to measuring the beam-induced $\gamma$-ray
			background and for beam tunings. The AFRODITE detection array mounted
			in a rhombicuboctahedron frame \cite{Sharpey-Schafer2004,Newman,Lipoglavek2006}
			served to us for detecting the nuclear $\gamma$-rays. It consisted
			of eight clover detectors of the EUROGAM phase II type \cite{Duchene}
			each containing four 50 x 70 mm HP-Ge crystals of n-type, housed in
			a common cryostat and Compton-suppressed by means of BGO crystals.
			In the configuration of fixed geometry used, four clovers were set
			at $\theta_{lab}$ = 90$^{{^\circ}}$ and the other four backward
			at 135$^{{^\circ}}$ with respect to the incident proton beam direction. Fig. \ref{fig:1} reports a schematic diagram of the experimental set up illustrating
			the passage of the proton beam across the AFRODITE reaction chamber
			equipped with the associated $\gamma$-ray clover detection array
			in the configuration used in our experiments.
			An average distance of $\approx$ 19.6 cm separated the front face
			of each HP-Ge crystal from the studied target. Then, the whole detection
			system subtended a solid angle amounting to a substantial fraction
			of $\approx$7.5$\%$ of the total $4\pi sr$. The individual HP-Ge
			crystals within a clover were set symmetrically at angles $\pm$5
			degrees relative to its center. This configuration of 32 HP-Ge crystals
			thus allowed us to measure $\gamma$-ray angular distributions with
			four detection angles each, i.e., $\theta_{lab}$ = 85$^{{^\circ}}$,
			95$^{{^\circ}}$ , 130$^{{^\circ}}$ and 140$^{{^\circ}}$. In the
			2016 experimental campaign we extended this limited angular range
			by placing one of the clovers at 169$^{{^\circ}}$, thus its HP-Ge
			crystals were positioned at $\theta_{lab}$ = 166$^{{^\circ}}$ and
			172$^{{^\circ}}$ (it had no BGO-Compton background suppression). 
                        This configuration of the AFRODITE $\gamma$-ray detection array actually
			allows for placing clovers at a forward angle relative to the incident
			beam direction, namely at $\theta_{lab}$ = 45$^{{^\circ}}$. However,
			given that the angular distributions of the $\gamma$-ray emission
			are expected to be symmetric relative to 90$^{{^\circ}}$ we voluntarily
			avoided this option as precaution for protecting the HP-Ge crystals
			against risky damages under high fluxes of neutrons produced in multi-scattering
			and reactions of the proton beam with the surrounding material elements
			(mainly Al and Ge) and the studied targets themselves. Both preceding
			TALYS code calculations and an experimental test run using a neutron
			detector performed prior to these experiments have pointed out the 
                        occurrence of important neutron backgrounds
			from these processes, especially at forward angles. The iron target
			was found to be the most critical one regarding neutron production
			via the $^{56}$Fe (p, xn) reaction. The corresponding $\sigma_{n}$
			total production section section is proton energy-dependent. E. g., at E$_{p}$
			= 66 MeV, we obtained a value of $\sim$1 b for this cross section
			that remains nearly constant over the proton energy range of E$_{p}$
			= 50 - 80 MeV, while lower values of $\sigma_{n}$ $\sim$ 150, 200
			and 250 mb were derived in case of the $^{12}$C, $^{24}$Mg and $^{28}$Si
			targets, respectively. Over the higher proton beam energy range of
			E$_{p}$ = 80 - 140 MeV, our Talys code calculations yielded a value of
			$\sim$350 mb for $\sigma_{n}$ around the characteristic neutron
			peak at energy E$_{n}$ = 1.0 MeV, then $\sigma_{n}$ (which varies as
			1/E$_{n}$) decreased to < 5 mb for higher neutron energies. Further
			details are provided on this point later in Subsection \ref{sub:transition}.
			
			We expected the observation of nuclear $\gamma$-ray lines of interest
			with energies extending up to about 7.2 MeV including the de-excitation
			lines at E$_{\gamma}$ = 6.129, 6.916 and 7.115 MeV from the 3$^{-}$
			, 2$^{+}$ and 1$^{-}$ low-lying states of $^{16}$O. We performed
			a careful energy calibration of the $\gamma$-ray detection array
			using standard radioactive sources of $^{60}$Co, $^{137}$Cs and
			$^{152}$Eu. The latter were placed at target position in the reaction
			chamber, and thus allowed us to cover the low photon energy range
			of E$_{\gamma}$ = 0.122 \textendash{} 1.408 MeV. Sometimes, we also
			used a locally produced $^{56}$Co source for extending the $\gamma$-ray
			energy range up to E$_{\gamma}$ $\approx$ 3 MeV. In addition, the
			prominent 6.129 MeV line of $^{16}$O and its escape lines produced
			in the (p, p\textquoteright $\gamma$) inelastic scattering off the
			Mylar target were used for photon energy calibration. The 
			energy resolution, frequently checked with respect to the 1.332 MeV
			line from the $^{60}$Co radioactive source, amounted typically to
			2.5 keV. We have used the same standard radioactive sources of $^{60}$Co,
			$^{137}$Cs and $^{152}$Eu for measuring the absolute efficiencies
			of the detection system over the low photon energy range up to E$_{\gamma}$
			= 1.408 MeV. Then, we extrapolated the obtained values to higher photon
			energies, up to E$_{\gamma}$ $\approx$ 10 MeV, via performing detailed
			Monte Carlo simulations of the AFRODITE detection array and surrounding
			material using the GEANT4 software \cite{GEANT4}.
			
			In each performed experimental run with the proton beam-on target
			under high vacuum, we recorded $\gamma$-ray energy spectra from the
			nuclear reactions of interest until the collection of good counting
			statistics in the less intense $\gamma$-ray peaks was obtained. We systematically
			recorded $\gamma$-ray background energy spectra from various related
			sources, such as experimental room, beam-induced, target and surrounding
			material activation backgrounds. Then, we used the accumulated background
			data for correcting the counts in the energy spectra of interest and
			as insights for possible spurious events in the latter. In particular,
			we subtracted the beam-induced backgrounds recorded with the incident
			protons passing through the Al empty-target frame from the measured
			main $\gamma$-ray energy spectra, upon normalizations to the same
			accumulated proton beam charges. Throughout all the experimental runs,
			we took great care for protecting the HP-Ge crystals via limiting
			their possible deterioration under the impact of excessive neutron
			backgrounds. The count rates in the detectors were not exceeding 5
			- 6 kHz per crystal, and were lower than 0.5 kHz during the runs with
			beam on the empty-target frame. All the experimental runs were in
			general operated with proton beam current intensities of at most 5-6
			nA and counting dead times of 6 - 8$\%$ (see also Ref. \cite{W.Yahia-Cherif2020}).
			
			During the experiments, we used a data acquisition system (the Multi
			Instance Data Acquisition System, MIDAS, \cite{mtsort}) based on
			digital electronics modules made by XIA of type \textquotedblleft DGF
			Pixie-16\textquotedblright.
			
			We worked the experimental runs generally in favorable, stable operating
			conditions, constantly monitoring the data acquisition system and
			the $\gamma$-ray count rates, and checking the levels of the neutron
			and $\gamma$-ray backgrounds in the recorded spectra. When unwanted
			variations occurred in the beforehand fixed conditions, we proceeded
			to additional tunings of the proton beam settings. We systematically saved the 
			recorded $\gamma$-ray energy spectra on discs together with timing and energy information.

			\section{Data analysis, experimental results\label{sec:Data-analysis,-experimental}}
			
			After the experiments, the count rate contents of the measured raw
			experimental data sets ($\gamma$-ray line energy and efficiency calibrations
			of the AFRODITE detection array, energy spectra and angular distribution data) have been checked
			and further corrected for dead time losses. Then, the data sets were
			analyzed using well-adapted homemade methods or software toolkits
			available in the literature as described below in this Section. 
                        In particular, the $\gamma$-ray line peaks within the energy spectra were
                        accurately identified with evaluating the spurious background lines, 
                        the relevant $\gamma$-ray line transitions were determined
			with carefully identifying the involved de-excitation nuclear states.
			Then, a line-shape analysis was performed with applying a special treatment
                        in order to extract the components contributed by different isotopes in observed, 
                        Doppler-broadened $\gamma$-ray line structures of complex shapes growing around
                        E$_{\gamma}$ = 4.44 MeV within the energy spectra from both two studied $^{nat}$C and Mylar targets. 
			
			Below, we first start with presenting a precise evaluation of the AFRODITE 
                        detection array absolute efficiencies using simultaneously experimental data from the $^{137}$Cs, $^{60}$Co
			and $^{152}$Eu calibrated radioactive sources up to E$_{\gamma}$
			= 1.408 MeV and Monte-Carlo GEANT4-simulated values over the photon
			energy range of 0.08 \textendash{} 10 MeV, thus extending the measured
			data up to E$_{\gamma}$ = 10 MeV. Finally, upon elaborated angular
			distribution Legendre polynomial least squares fits to the $\gamma$-ray
			differential experimental cross section data, we determine the corresponding
			angle-integrated $\gamma$-ray line production cross sections. Detailed
			accounts of the uncertainties affecting each of the above quantities
			are provided below in this section.

			\subsection{$\gamma$-ray detection efficiencies\label{sub:sub-section-A}}
			
			As indicated in the previous section, we have evaluated the absolute
			$\gamma$-ray detection efficiency, $\varepsilon(E_{\gamma},\theta)$,
			for a large photon energy domain of interest. Experimental values
			taken with the calibrated radioactive sources covered the energy range
			of $E_{\gamma}$ = 122 \textendash{} 1408 keV. They were determined
			using the relation
			
			\begin{equation}
			\varepsilon(E_{\gamma})=\frac{N_{\gamma}(E_{\gamma})}{A(t{)}\times\Delta t\times I},\label{eq:eff1}
			\end{equation}
			in terms of the $\gamma$-ray line count integrals (photo-peak areas)
			in the $\gamma$-ray energy spectra, $N_{\gamma}(E_{\gamma})$, the
			radioactive source activities at the measurement time, $A(t)$,
			the acquisition time interval, $\Delta t$, and the nuclear $\gamma$-ray
			branching ratios, $I$. 
			
			The uncertainty in the detection efficiencies obtained with the radioactive
			sources is dominated by the uncertainty in the source activity amounting
			to 3.7 \textendash{} 5 \%, while the branching ratios contribute by
			less than 1\% and the statistical error in the peak areas amounts to 1-2\%.
			
			We extended the $\gamma$-ray energy range up to 3.5 MeV by using
			efficiency data from a previous experiment \cite{Lipoglavek2006} carried out 
            with an uncalibrated $^{56}$Co source under identical conditions as in the 
			present work, normalized to our experimental data. The systematic uncertainty 
			in this normalization was found to be of less than 2\% based on the deviation 
			in the region of overlap between the data sets.
			
			Extrapolation of the $\varepsilon(E_{\gamma},\theta)$ data to higher
			photon energies was based on Monte-Carlo simulations. 
			Indeed, simulating the $\gamma$-ray emission from the radioactive sources using the GEANT4 program \cite{GEANT4}, we first reproduced the corresponding experimental spectra as shown in Fig. \ref{fig:2}. 
			Then, simulated efficiency values were determined for each HP-Ge crystal, 
			allowing us to extend the data from the sources to higher $\gamma$-ray energies. Gamma-ray line sum peaks following true coincidence-cascades between full-absorption energy $\gamma$-rays from the sources, expected to
			grow in the experimental spectra, were found to be of very low intensities
			amounting to less than 0.05$\%$ relative to the intensities of the
			latter photo-peak lines. Consequently, they should not have appreciable
			effects on the absolute detection efficiency data.
			
			While the GEANT4 simulations tend to overestimate the absolute efficiency,
			it had been shown that they reproduce the energy dependence of the
			HP-Ge detection efficiencies accurately, with no adjustable parameters
			\cite{Belhout2007,Benhabiles-Mezhoud2010,W.Yahia-Cherif2020}. The GEANT4 
			simulations were performed at source energies and at regular energy intervals 
			between 80 keV and 10 MeV. They were normalized to the efficiencies obtained with the radioactive
			sources over the photon energy range of E$_{\gamma}$ = 0.5 - 1.4
			MeV, the scaling factors between the two data sets amounting to 0.87, in  average. 
			
			We then derived the values of $\varepsilon(E_{\gamma},\theta)$ for
			each detector and versus the photon energy via interpolating all the data
			using the following multi-parameter function reported in the RadWare software
			package \cite{Radford}, i.e.,

			\begin{equation}
			\epsilon(E_{\gamma},\theta)=e^{[(A+BX+CX^{2})^{-G}+(D+EY+FY^{2}%
			)^{-G}]^{-1/G}}, \label{eq.2}%
			\end{equation}
                        where $X=log(E_{\gamma}/100)$, $Y=log(E_{\gamma}/1000)$ and the other quantities are adjustable free parameters.

			The fitted curves were used for deriving the absolute detection efficiency values for each observation angle and versus the photon energy, and subsequently for calculating the $\gamma$-ray cross sections.

		       Fig. \ref{fig:3} shows the measured efficiencies, the normalized GEANT4 simulated values and the fitted curve to these data points (solid line). The dotted lines in this figure indicate an average uncertainty of the order of 12\% estimated here for the overall absolute efficiencies of the AFRODITE detection array. For this purpose, we mainly
                       took into account the statistical errors resulting from the calculation of the photopeak areas
		       within the $\gamma$-ray energy spectra, to which we added in quadrature the uncertainties
		       in the radioactive source activities, in the $\gamma$-ray branching ratios, in the GEANT4-simulated 
		       values, in the normalization factors and in the fits of the multi-parameter function to the data. 

		       The values of the absolute efficiencies at high energies are further supported by the
                       obtained good agreement between our experimental cross sections reported here for the lines 
                       at E$_{\gamma}$ = 4.44 and 6.129-MeV of $^{12}$C and $^{16}$O, respectively,
                       and the previously measured values (available for beam energies below 50 MeV). 
			
			\subsection{$\gamma$-ray energy spectra, transitions properties \label{sub:transition}}
			
			One can use two operating modes for treating the raw experimental
			data ($\gamma$-ray count rates, energy spectra, detection efficiencies,
			etc.) from a given clover, either the single HP-Ge crystal mode or
			the add-back mode consisting in summing the energies detected in two
			adjacent crystals of the clover. In this work we used the former mode in the data
			analysis given that the counting statistics from each HP-Ge crystal
			was sufficient, thus, we considered each Ge crystal as an individual
			Compton background-suppressed detector. To improve the Compton 
			background suppression we also rejected coincident events between 
			elements of a clover detector.  Along this phase, we often checked the 
			consistency of the estimated experimental values of $\gamma$-ray cross 
			sections with corresponding orders of magnitude predicted by nuclear reaction 
			models using the TALYS code.
			
			After the experiments, we continued the treatment and analysis of
			the recorded experimental data in order to extract as accurately as
			possible $\gamma$-ray angular distribution and production cross section
			data. We analyzed the $\gamma$-ray peaks within the photon energy
			spectra following the method detailed in Refs. \cite{W.Yahia-Cherif2020,Belhout2007,Belhout2009}
			by means of the ROOT software \cite{ROOT} or the gf3 software \cite{Radford}.
			We extracted their areas by fitting symmetric Gaussian-shape distributions
			to the measured data with subtracting the related backgrounds by linear
			function fits. However, Gaussian shape fits to the experimental data
			were not always applicable, and a special method was used to extract
			the $\gamma$-ray line areas as, for instance, for the line of $^{12}$C
			at E$_{\gamma}$ = 4.439 MeV (see the next subsection). We evaluated
			the total relative uncertainty of each peak area as the sum in quadrature
			of the uncertainty resulting from the fitting procedure, the statistical
			uncertainty and the systematic uncertainty. The latter two uncertainties
			amounted to at most 5 and 10$\%$, respectively. As described in the
			previous section, we performed precise determinations of the $\gamma$-ray
			line energies within the recorded spectra with absolute uncertainties
			typically amounting to less than 0.1 keV. Fig. \ref{fig:4}
			reports examples of experimental energy spectra in single mode from
			two individual HP-Ge detectors located, respectively, at observation
			angles, $\theta_{lab}$ = 130$^{{^\circ}}$ and 85$^{{^\circ}}$:
			(a) for a 54 MeV proton beam incident on the $^{nat}$C target and
			(b) in case of a 150 MeV beam hitting the Mylar foil. One can notice 
			great similarities in the general shapes ($\gamma$-ray lines + backgrounds) 
			of these $\gamma$-ray energy spectra with presence of the intense background 
			line at 511 keV from the positron-electron pair annihilation, as well as marked differences
			featuring their contents in $\gamma$-ray line peaks due to the differences
			in their respective target compositions. Fig. \ref{fig:4} also illustrates
			that while the Compton background was reduced by the BGO suppression,
			it was not fully removed. Figs. \ref{fig:5} and \ref{fig:6} report
			separately the $\gamma$-ray energy spectra from the same two HP-Ge
			detectors, corrected for proton beam-induced backgrounds, after further
			subtracting the Compton background and the background from the reaction
			chamber in the subsequent data analysis phase.
			
			The recorded $\gamma$-ray energy spectra thus exhibit several $\gamma$-ray
			lines of various shapes and intensities, the most intense, narrow
			peak of interest corresponds to the line of $^{10}$B at E$_{\gamma}$
			= 0.718 MeV (see Figs. \ref{fig:5} and \ref{fig:6}). Some of the
			observed lines, like the lines of $^{11}$C and $^{12}$C at E$_{\gamma}$
			= 2.000 and 4.439 MeV, respectively, are subject to strong Doppler
			broadening and shift making their analysis rather difficult, as will
			be clarified in next subsection. We pointed out 20 peaks corresponding
			to de-excitation lines from various isotopes produced in proton-induced
			reactions. Notably, the lines at E$_{\gamma}$ = 4.439 and 6.129 MeV referring to the E2 (2$^{+}$, 4.439 MeV, $\tau$ = 42 fs $\rightarrow$
			0$^{+}$, g.s) and E3 (3$^{-}$, 6.129 MeV, $\tau$ = 18.4 ps$\rightarrow$
			0$^{+}$, g.s) transitions in $^{12}$C and $^{16}$O, respectively,
			produced in (p, p\textquoteright $\gamma$) inelastic proton scattering,
			suffer considerable Doppler broadenings (\ensuremath{\approx} 150
			keV for the 6.129 MeV line), which makes them good candidates for
			line shape analyses \cite{H.Benhabiles-Mezhoud2011,Kiener2019}. Tables
			(\ref{tab:2}, \ref{tab:3}) display separately the lists and properties
			of the identified $\gamma$-ray transitions resulting from proton-beam-induced reactions on the $^{nat}$C and Mylar target, respectively.
			They correspond to the de-excitation of various low-mass isotopes
			($^{10,11}$B, $^{11,12,13}$C, $^{14,15}$N, $^{15,16}$O) produced
			in (p, p\textquoteright $\gamma$) inelastic proton scattering, fusion-evaporation
			reactions and spallation reactions induced by high-energy protons
			on $^{12}$C and $^{16}$O (see Tables (\ref{tab:2}, \ref{tab:3})).
			As can be seen in these tables, the (E$\lambda$, M$\lambda$) line
			transitions pointed out have essentially M1, E2, E3, (M1 + E2),
			(M2 + E3) and (E1 + M2 + E3) characters with multipolarity values
			not exceeding $\lambda$ = 3. In addition, the energy spectra show
			the presence of background peaks for $\gamma$-ray lines emitted in
			proton and secondary neutron-induced reactions on the nuclei of surrounding
			Al and Ge elements from the reaction chamber, the target frames and
			holder, and the HP-Ge detectors \cite{Gilmore,Bunting1974}. In particular,
			several $\gamma$-ray lines with asymmetric shapes appear in the energy
			spectra. They refer to background de-excitation lines produced in
			inelastic secondary neutron scattering off Ge isotopes. These are
			emission lines at E$_{\gamma}$ = 198 keV coming from the $^{70}$Ge(n,
			$\gamma$)$^{71m}$Ge radiative neutron capture reaction, E$_{\gamma}$
			= 595.85 keV from the (n, n\textquoteright $\gamma$) inelastic neutron
			scattering of $^{74}$Ge, and E$_{\gamma}$ = 689.60, 834.01 and 894
			keV from the (n, n\textquoteright $\gamma$) inelastic neutron scattering
			of $^{72}$Ge. In addition to the 511 keV line, the energy spectra
			contained background lines from $^{27}$Al, like the lines at E$_{\gamma}$
			= 843.76 keV resulting from both the $^{27}$Al(n, p) $^{27}$Mg$^{\ast}$($\beta$$^{-}$)$^{27}$Al
			sequential process and the (p, p\textquoteright $\gamma$) inelastic
			proton scattering of $^{27}$Al, 1.014 MeV from the latter reaction
			and 1.369 MeV from the $^{27}$Al(p, $\alpha\gamma$)$^{24}$Mg reaction,
			etc. Notice that apart from the 198 keV background line, no $\gamma$-ray
			lines of interest dominated the low-energy part of the $\gamma$-ray
			energy spectra below the positron-electron annihilation line at E$_{\gamma}$
			= 511 keV. Table \ref{tab:4} lists most of the observed background
			lines, together with the corresponding nuclear reactions. The background
			lines practically affected all the $\gamma$-ray energy spectra recorded
			in our experiments involving the proton beams (see, e.g., Figs. \ref{fig:4}
			(a, b)).

			\subsection{$\gamma$-ray line shape data, analysis of the 4.44 MeV line complexes\label{sub:-ray-line-shape}}
			
			As one can see in Figs. \ref{fig:4} (a, b), a broad structure appears
			at E$_{\gamma}$ $\sim$ 3.3 - 4.5 MeV in the $\gamma$-ray energy
			spectra from both natural C and Mylar targets. These are actually
			line complexes composed of overlapping, Doppler-broadened lines of
			interest, together with their Compton components and escape peaks.
			In the case of the C target, in addition to the 4.439-MeV line, which
			is the dominant line, two other $\gamma$-ray lines at E$_{\gamma}$
			= 4.319 and 4.444 MeV attributed to $^{11}$C and $^{11}$B, respectively,
			are significant components of the complex (see Fig. \ref{fig:5},
			Table \ref{tab:3}). Nuclear reaction code calculations predict that
			the 4.439 MeV line of $^{12}$C is the only component below E$_{p}$
			= 25 MeV, while above this energy the 4.445 MeV line of $^{11}$B
			significantly contributes via the $^{12}$C(p, 2p)$^{11}$B reaction,
			as well as the 4.319 MeV line of $^{11}$C via the $^{12}$C(p, pn)$^{11}$C
			reaction. Some minor components like the 4.339 MeV line of $^{11}$C
			\cite{Kiener2019} and the 4.444 MeV line of $^{10}$B are probably
			also present.
			
			In such conditions, the determination of the full-energy peak contents
			requires a special treatment. Therefore, we were led to perform meticulous
			spectral deconvolution concerning the peak associated with the line
			of $^{11}$C at E$_{\gamma}$ = 4.319 MeV appearing on the left side
			of the line complex at E$_{\gamma}$ = 4.44 MeV. To this end, we generated
			the corresponding theoretical spectral shape as follows. We subtracted
			the Compton background and escape peaks by means of the $\gamma$-ray
			energy-deposit spectra in the HP-Ge crystals that were simulated by
			the GEANT4 program over an extended photon energy domain, 0.5 $\leq$
			E$_{\gamma}$ $\leq$ 7 MeV. The response functions of the HP-Ge detectors
			also required careful checking via simulations of the $\gamma$-ray
			emission spectra, which one could do for standard radioactive sources
			of known branching ratios placed at the target position and then compare
			the simulated and experimental spectra. This is illustrated in Fig. \ref{fig:2} (discussed in Subsection \ref{sub:sub-section-A}.) reporting
			an example of experimental spectra from calibrated $^{152}$Eu and
			$^{60}$Co radioactive sources recorded with an HP-Ge detector in
			the 2016 experimental campaign together with their GEANT4-simulated
			counterparts. The observed very good agreement in this
			figure between the experimental and simulated energy spectra then
			attests for the correctness of the operating methods used in the current
			work.
			
			We derived the contribution of the 4.319 MeV line of $^{11}$C after
			calculation of its line shape following the method described in Ref.
			\cite{Kiener2001}. For this purpose, we used a simulation program
			of Monte-Carlo type written by Kiener \cite{Kienercode} with taking
			into account the geometry and the energy resolution of the HP-Ge detector.
			We used as input data for the $^{12}$C(p, p n)$^{11}$C reaction
			in this program the angular distributions for particle emission and
			the mean excitation energy of the nuclear state derived by TALYS code
			calculations. We assumed the slowing down within the target of the
			$^{11}$C isotope in its excited state ($\tau$ < 8.3 fs) to occur
			with exponentially decreasing $\gamma$-ray emission probability,
			evaluating the target stopping power by the SRIM code \cite{SRIM}.
			
			Fig. \ref{fig:7} reports the obtained results for the simulated spectra
			for the $^{11}$C line at E$_{\gamma}$ = 4.319 MeV, while Fig. \ref{fig:8}
			illustrates the evolution versus the proton beam energy and observation
			angle of the line shape data in the complex at E$_{\gamma}$ = 4.44
			MeV after subtraction of the simulated spectra for the line of $^{11}$C.
			Obviously, the separation of the Doppler-broadened peaks at E$_{\gamma}$
			= 4.439 and 4.444 MeV for the lines of $^{12}$C and $^{11}$B, respectively,
			is impossible. We therefore attributed the counts remaining after
			the subtraction of the 4319-keV line to a line labelled 4.44-MeV,
			which is the sum of the 4439-keV, 4444-keV and some minor lines as
			discussed above.
			
			The interaction of the proton beam with the Mylar (C$_{10}$H$_{8}$O$_{4}$)
			target produces nuclear reactions simultaneously with $^{12}$C and
			$^{16}$O, which leads to the superposition of de-excitation $\gamma$-ray
			lines from these two target nuclei. In this case also, the extraction
			of the areas under the peaks for the line of $^{12}$C at E$_{\gamma}$
			= 4.439 MeV produced in the $^{16}$O(p, p\textquoteright $\alpha$)$^{12}$C
			reaction required a special treatment.
			
			After subtraction of the Compton backgrounds and the 4.319 MeV line
			of $^{11}$C similarly as explained above, the 4.44 MeV complex in
			each $\gamma$-ray energy spectrum contained four components. Two
			components for the 4.439 line of $^{12}$C resulting from the $^{12}$C(p,
			p\textquoteright )$^{12}$C and $^{16}$O(p, p\textquoteright $\alpha$)$^{12}$C
			reactions, respectively, and the two components for the line of $^{11}$B
			at E$_{\gamma}$ = 4.444 MeV from reactions with $^{12}$C and $^{16}$O.
			In the spectra from this target (Mylar), the count integrals (N$_{\gamma}^{tot}$)
			are sums of the integrals (N$_{\gamma}^{p}$, N$_{\gamma}^{\alpha}$
			) from the two preceding reactions, respectively.
			
			We subtracted the 4.439 MeV and 4.444 MeV line components (N$_{\gamma}^{p}$)
			due to reactions with $^{12}$C with the help of the energy spectra
			recorded in proton irradiations of the natural C target ($\gamma$-ray
			line integrals, N$_{\gamma}^{c}$), following the respective accumulated
			beam charges in the runs with the C and Mylar targets and the respective
			numbers of C atoms in each target.
			
			As in the preceding case (lines in the spectra from the natural C
			target), the two components at 4.439 MeV and 4.444 MeV (from $^{12}$C
			and $^{11}$B, respectively) of the line complex in each energy spectrum
			from the $^{16}$O isotope in the Mylar target are attributed to the
			4.44-MeV line.

			\subsection{$\gamma$-ray angular distributions and integral cross sections}
			\subsubsection{Angular distributions}
			
			We derived the $\gamma$-ray differential experimental cross section
			data from the extracted $\gamma$-ray line count integrals, $N_{\gamma}(E_{\gamma},\theta)$,
			the absolute detection efficiencies $\varepsilon(E_{\gamma},\theta)$,
			the target thicknesses (number of nuclei per cm${{}^2}$), $N_{t}$,
			and the numbers of projectiles falling on the targets, $\phi$, (recorded
			in the experimental runs) using the relation
			\begin{equation}
			\frac{d\sigma(E_{\gamma},\theta)}{d\Omega}=\frac{N_{\gamma}(E_{\gamma},\theta)}{\varepsilon(E_{\gamma},\theta)N_{t}\phi}10^{27}(mb/sr).		
			\label{eq:1}
			\end{equation}
			
			We then fitted the following angular distribution Legendre polynomial
			expansion to the measured data, i.e.,
			\begin{equation}
			W(\theta)=\sum_{\ell=0}^{\ell_{max}}a_{\ell}Q_{\ell}P_{\ell}\left(\cos(\theta)\right).\label{eq:2}
			\end{equation}

			The summation in this expression extends over only even, integer $\ell$-values,
			with $\ell_{max}$ taking twice the $\gamma$-ray multipolarity
			or twice the spin of the emitting state. Since the studied $\gamma$-ray
			lines are featured by at most $\lambda$ = 3 (see Tables (\ref{tab:2},
			\ref{tab:3})), this then fixes $\ell_{max}$ = 6. The Q$_{\ell}$
			are energy-dependent geometrical attenuation coefficients \cite{Rose1953,iliadis2007nuclear},
			functions of ($\varepsilon$, $\theta$, $\ell$), that were introduced
			for taking into account the finite dimensions of the $\gamma$-ray
			detectors and for better evaluating the angular distribution expansion.
			Their precise determination via GEANT4 Monte-Carlo simulations of
			the AFRODITE detection array yielded roughly constant values equal
			or close to unity over the photon energy range of interest, 0.1 MeV
			$\leq$ E$_{\gamma}$ $\leq$ 10 MeV, with relative uncertainties
			of $\sim$1$\%$. Notice that since one is dealing with proton-induced
			reactions on appreciably heavier target nuclei, all the results reported
			here were derived under the assumption, $\theta=\theta_{lab}\approx\theta_{cm}$.
			We report as examples in Fig. \ref{fig:9} (a, b) the experimental
			angular distribution data derived versus the proton energy for the main,
			prominent $\gamma$-ray lines of $^{12}$C and $^{16}$O at 4.44
			and 6.129 MeV, respectively, together with corresponding least-squares
			Legendre polynomial best-fit curves generated by Eq. (\ref{eq:2}).
			Table \ref{tab:5} displays the associated $a_{\ell}$ fit coefficients.
			The reduced $\chi^{2}$ values resulting from the fits amounted to
			one or below one for the narrow, isolated $\gamma$-ray lines, and
			to at most three in case of overlapping, broad line structures in
			the energy spectra. We derived the uncertainties in the experimental
			differential cross sections by adding in quadrature the errors in
			the quantities used in Eq. (\ref{eq:1}). An overall average value
			of $\sim$17$\%$. of the relative uncertainty in these data then
			resulted.
			\subsubsection{Integral cross sections and associated uncertainties \label{sub:-ray-line-production}}
			The angle-integrated $\gamma$-ray line production cross sections
			were determined from the a$_{0}$ coefficients of the angular distribution
			expansions given by Eq. (\ref{eq:2}) using the relation $\sigma=4\pi a_{0}$.
			Among 20 identified $\gamma$-ray lines of interest in the recorded
			energy spectra (9 lines from the $^{nat}$C target and 11 lines from
			Mylar, see Tables (\ref{tab:2}, \ref{tab:3})), only 10 lines (5 lines from each target)
			were found intense enough to extract integrated cross sections. We
			report in Figs. (\ref{fig:10}, \ref{fig:11}) the obtained $\sigma(E_{p})$ excitation function
			results for the analyzed most intense five $\gamma$-ray lines observed
			in the proton irradiations of the $^{nat}$C target. We also report
			in Figs. (\ref{fig:12}, \ref{fig:13}) similar results obtained for the strongest five
			lines from the proton irradiations of the Mylar target. Finally, we
			list in Table \ref{tab:6} the determined values of the integrated $\gamma$-ray
			line cross sections for the identified ten most intense $\gamma$-ray
			lines emitted in proton irradiations of the $^{nat}$C and Mylar targets.
			Notice that only five experimental data points (from expected 11)
			are present in the excitation function for the 6.129 MeV octupole
			$\gamma$-ray transition in $^{16}$O. These data points come from
			the runs of the 2016 experimental campaign performed using the extended
			detection setup with the HP-Ge detector placed at $\theta$ = 169$^{{^\circ}}$.
			
			The error bars in the differential cross sections 
			influence  the Legendre polynomial expansion (Eq.
			\ref{eq:2}) fits to the data leading to the values of the a$_{\ell}$ coefficients
			and associated errors, in particular for the a$_{0}$ term enabling
			the calculation of the integral cross sections and related errors.
			
			We have primarily estimated the errors
			propagating in Eq \ref{eq:1} that affect the quantities used for calculating
			the experimental differential cross sections.
                        Statistical errors were first considered alone in the fit procedure,
			then systematic errors were added in quadrature in order to generate
			the final values of the uncertainties in the determined $\sigma$$(E_{p})$
			data. The systematic uncertainties usually affect equally the measurements
			at all the observation angles. We have mainly taken into account the
			errors in the target thicknesses (1$\%$), the errors associated with
			the absolute detection efficiencies, $\varepsilon(E_{\gamma},\theta)$,
			typically amounting to at most 12$\%$, as discussed in Subsection \ref{sub:sub-section-A}
			and those in the collected beam charge (1$\%$). As statistical uncertainties,
			we considered the errors in the $\gamma$-ray line count integrals
			(line peak areas), $N_{\gamma}(E_{\gamma},\theta)$, within the measured
			$\gamma$-ray energy spectra. Typically, these errors do not exceed
			5$\%$ in case of isolated, narrow $\gamma$-ray lines of Gaussian shapes.
			In contrast, for Doppler-broadened, overlapped lines of non-Gaussian
			shapes like the line complexes at E$_{\gamma}$ = 4.44 MeV (see Subsection
			\ref{sub:-ray-line-shape}), the evaluation of statistical errors is much more complicated,
			yielding much larger values of at least 10-12$\%$.

			In summary, adding in quadrature
			the above two types of uncertainties (systematic and statistical)
			and the uncertainty in the Legendre polynomial fits, we obtained
			mean values of 18-25\% for the overall relative uncertainty in the
			$\sigma(E_{p})$ data in case of isolated, narrow lines, and values
			in the range of 20 - 30\% in case of Doppler-broadened, overlapping
			lines. These uncertainties are also depicted in Figs. (\ref{fig:10}-\ref{fig:13}) and listed
			in Table \ref{tab:6} together with the corresponding values of the integral
			$\gamma$-ray line production cross sections.

			\section{Comparison of our $\gamma$-ray production cross sections to previous
				data sets\label{sec:Comparison-of-our}}
			
			One remarks in Figs (\ref{fig:10}-\ref{fig:13}) that among the analyzed
			10 intense lines from the two targets, previous experimental data
			sets are mostly available for the two main lines of $^{12}$C and
			$^{16}$O at E$_{\gamma}$= 4.439 and 6.129 MeV, respectively, the
			former line being produced in proton irradiations of both the $^{nat}$C
			and Mylar targets (see Tables (\ref{tab:2}, \ref{tab:3}). They cover
			especially the low proton energy region, E$_{p}$ $\leq$ 26 MeV,
			dominated by the compound resonance structure at E$_{p}$ = 20 \textendash{}
			30 MeV investigated with tandem accelerators by the Orsay \cite{H.Benhabiles-Mezhoud2011,Belhout2007,
			Belhout2009,Benhabiles-Mezhoud2010,Kiener2008,Kiener1998}
			and Washington \cite{Dyer81,Narayanaswamy81,Seamster84,Dyer85} groups,
			while only a few data points taken with cyclotrons \cite{Lesko1988,Lang1987}
			are present at higher proton energies.

			\subsection{Lines from the $^{nat}$C target}
			
			We have identified nine lines in proton irradiations of the $^{nat}$C
			target (see Table \ref{tab:2}). We extracted the $\sigma(E_{p})$
			data shown in Figs. (\ref{fig:10}, \ref{fig:11}) for the five most
			intense lines. These are the main, prominent line at E$_{\gamma}$
			= 4.439 MeV from inelastic proton scattering of $^{12}$C (set at
			E$_{\gamma}$ = 4.44 MeV as explained in subsection \ref{sub:-ray-line-shape}),
			and the lines at E$_{\gamma}$ = 0.718 and 1.022 MeV of $^{10}$B,
			E$_{\gamma}$ = 2.000 MeV of $^{11}$C and E$_{\gamma}$ = 2.124 MeV
			of $^{11}$B produced in various other nuclear reactions (see Table
			\ref{tab:2}). Regarding the 4.439 MeV main line of $^{12}$C, one
			can observe in Fig. \ref{fig:10} that our experimental data are in
			very good agreement with practically all the previous data sets measured
			at cyclotron facilities by Lesko et al. \cite{Lesko1988} and Lang
			et al. \cite{Lang1987} over the common proton energy range of E$_{p}$
			= 30 - 85 MeV. They also show to be fully consistent with the lower
			energy data sets taken previously at tandem accelerators below E$_{p}$
			= 30 MeV \cite{H.Benhabiles-Mezhoud2011,Dyer81,Belhout2007,Belhout2009,Benhabiles-Mezhoud2010,Kiener2008,Kiener1998}.
			
			Experimental data have been measured previously by Lang et al. \cite{Lang1987}
			at three proton energies, E$_{p}$ = 40, 65 and 85 MeV for the lines
			of $^{11}$C and $^{11}$B at E$_{\gamma}$ = 2.000 MeV and E$_{\gamma}$
			= 2.124 MeV, respectively. As one can see in Fig. \ref{fig:11}, their
			data at E$_{p}$ = 40 MeV agree very well with our data at E$_{p}$
			= 42 MeV, while their two other values lie slightly higher relative
			to our values, mainly for the line of $^{11}$B at E$_{\gamma}$ =
			2.124 MeV. Finally, concerning the two lines of $^{10}$B at E$_{\gamma}$
			= 0.718 and 1.022 MeV, no previous experimental values were reported
			in the literature that could be compared to our data points (see Fig.
			\ref{fig:11}). While a second cross-section maximum around 70 MeV
			due to fusion-evaporation reactions, predicted by TALYS \cite{TALYS}
			and the compilation of Murphy et al. \cite{R.J.MurphyB.KozlovskyJ.Kiener2009},
			has been indeed pointed out in the recent experiments \cite{Kiener2021PhyRevC}
			with accelerated $\alpha$-particles on C, Mg, Si and Fe targets,
			no evidence for similar trend is shown by the present experimental
			data from proton-induced reactions.

			\subsection{Lines from the Mylar target}
			
			We report in Figs. (\ref{fig:12}, \ref{fig:13}) experimental $\sigma(E_{p})$
			data for the five most intense lines among eleven lines identified
			in the proton irradiations of the Mylar target (see Table \ref{tab:3}).
			One can see in Fig. \ref{fig:12} that for the octupole line of $^{16}$O
			at E$_{\gamma}$ = 6.129 MeV only five experimental data points are
			shown at E$_{p}$ = 66, 125, 150, 175 and 200 MeV out of the 11 proton
			energies explored in our experiments, due to the angular distribution
			restriction, as stated. Therefore, the comparison to previous data
			sets is also limited in this case. However, one observes that the
			data point of Lang et al. at E$_{p}$ = 65 MeV appears to match excellently
			our $\sigma$-value at 66 MeV. In contrast, the data points taken over the
			proton energy range of E$_{p}$ = 23 - 50 MeV by Narayanaswamy et
			al. \cite{Narayanaswamy81} (E$_{p}$ = 23.7 and 44.6 MeV), Lesko
			et al. \cite{Lesko1988} (E$_{p}$ = 30, 32, 40 and 50 MeV), and Lang
			et al. \cite{Lang1987} (E$_{p}$ = 40 and 85 MeV) seem to overestimate
			the trend of our experimental values. The latter are well reproduced
			by nuclear reaction models via our TALYS calculations using modified
			OMP parameters (see next section). Regarding the 4.439 MeV line of
			$^{12}$C produced via the $^{16}$O(p, p\textquoteright $\alpha$)$^{12}$C
			reaction (see Fig. 13 (a)) our data set clearly appears to be in very good agreement
			within the experimental uncertainties with the previous
			data set of Lesko et al. \cite{Lesko1988} and the value of Lang et
			al. \cite{Lang1987} over the explored common proton energy range,
			E$_{p}$ = 30 \textendash{} 50 MeV. They are also very consistent
			with the previous experimental data set from tandem accelerators below
			E$_{p}$ = 26 MeV. For the two lines of $^{15}$O and $^{15}$N at
			E$_{\gamma}$ = 5.240 and 5.269 MeV, respectively, previous data sets
			were measured at E$_{p}$ = 30 - 40 MeV by Lesko et al. \cite{Lesko1988}
			and at E$_{p}$ = 40 - 85 MeV by Lang et al. \cite{Lang1987}, while
			for the 2313 keV line of $^{14}$N only data from the latter authors
			exist. As one can observe in Fig. \ref{fig:13}, the previously reported
			values for these lines show a general agreement with our cross section
			curves with, however, some disagreement in the absolute values, in
			particular for the data of Lang et al. \cite{Lang1987}.
			
			One can finally conclude that our experimental excitation functions
			for all the analyzed ten lines agree quite well, within the
			involved experimental uncertainties, with the previously reported
			cross section data sets, extending coherently the latter to higher
			proton energies up to E$_{p}$ = 200 MeV.

			\section{Comparison of experimental $\gamma$-ray production cross sections
				to nuclear reaction models and to the Murphy et al. compilation, discussion.\label{sec:Comparison-of-experimental}}

			\subsection{TALYS code calculations, comparison to nuclear reaction model predictions}
			
			The nuclear reaction code TALYS \cite{TALYS} proved to be a very
			powerful tool for interpreting laboratory measured nuclear data such
			as nuclear reaction cross sections for fundamental research and various
			applications. It reliably predicts their values at energies out of
			reach to particle accelerators or where experimental data were missed
			\cite{R.J.MurphyB.KozlovskyJ.Kiener2009}. EMPIRE2 \cite{M.HermanEMPIRE}
			or TALYS were used for establishing nuclear data libraries \cite{koning2019tendl,sallaska2013starlib},
			for predicting nuclear reaction rates and related uncertainties, and
			chemical abundances in astrophysical media \cite{sallaska2013starlib,Goriely2008,Harissopulos}.
			Several groups extensively used TALYS for reproducing $\gamma$-ray
			experimental cross sections \cite{kozlovsky2002nuclear,R.J.MurphyB.KozlovskyJ.Kiener2009,H.Benhabiles-Mezhoud2011,W.Yahia-Cherif2020,Lesko1988,Kiener2021PhyRevC}
			applied for modeling solar flares, the interactions of galactic LECRs,
			and calculating the total $\gamma$-ray emission fluxes from these
			sites \cite{benhabiles2013erratum,T.VieuS.GabiciV.Tatischeff,kozlovsky2002nuclear,W.Yahia-Cherif2020}.
			Tatischeff, Kozlovsky, Kiener and Murphy \cite{Tatischeff2006} used
			both codes for calculating the delayed $\gamma$-ray line emission
			by radioactive ions produced in solar flares.
			
			Cross sections can be calculated by TALYS for nuclear reactions induced
			by $\gamma$-ray, neutrons and charged particles (p, d, t, $^{3}$He
			and $^{4}$He) on various target nuclei over the energy range of E$_{lab}$
			= 0.001 - 250 MeV, based on main reaction mechanisms expected in this
			region (compound nucleus, direct reactions, pre-equilibrium, fusion-evaporation,
			etc.). In addition, the code includes libraries of input nuclear data
			(masses, level densities, discrete states, OMP and $\beta_{\lambda}$
			level deformation parameters, $\gamma$-ray branching ratios, strength
			functions. etc.) but allows the user to use instead values of these
			parameters either determined experimentally or derived by theoretical
			models. The user could then perform calculations with the default
			input parameters of TALYS or alternatively input one\textquoteright s
			own values (labelled modified parameters of TALYS) by means of provided
			keywords (in form of input files) or by editing the intrinsic nuclear
			data libraries of the code. We used the 1.6 version of TALYS, where
			experimentally known nuclear states are supposed to decay exclusively
			via $\gamma$-ray electromagnetic transitions. This version is suitable
			in the case of medium and heavy nuclei, where usually the first 25
			states (the maximum number of states in Talys, taken from experimental
			tables) are stable against particle emission and decay exclusively
			by electromagnetic transitions. Otherwise, for light target nuclei
			like $^{12}$C, for example, which has only one excited state below
			the alpha emission threshold, the user should modify the $\gamma$-ray
			branching ratios to take account of particle emission, or reduce the
			number of experimentally known states, explicitly used in the Talys
			calculations.
			
			We have first started calculating $\gamma$-ray production cross sections
			for the analyzed strongest lines of Tables (\ref{tab:2}, \ref{tab:3})
			using the default input parameters of TALYS, introducing only the
			incident energy values and the (Z, A) properties of the projectile
			and targets. We obtained results that reproduced only the low-energy
			part of the experimental cross section data but considerably deviated
			from them elsewhere over wide energy domains, which led us to modify
			some default input data in TALYS. The required modifications concerned
			mainly the $\beta_{\lambda}$ level deformation parameters for $^{12}$C,
			and the optical model potential parameters for both $^{12}$C and
			$^{16}$O target nuclei. In the case of the $^{12}$C target, we adjusted
			the OMP parameters such that the calculated cross sections reproduced
			the experimental data for the main line of $^{12}$C at E$_{\gamma}$
			= 4.44 MeV. The obtained modified OMP values (stored in an input file)
			were close to those extracted from the analyses \cite{Kato1985,Fannon1967,Meigooni1985,Perey1976}
			of angular distributions for inelastic proton scattering, while we
			have taken the $\beta_{\lambda}$ deformation parameters ($\beta_{2}$=
			- 0.61, $\beta_{4}$= 0.05) from Ref. \cite{Meigooni1985}. We proceeded
			similarly in order to reproduce the main line of $^{16}$O at E$_{\gamma}$
			= 6.129 MeV. Notice that in the calculation of the theoretical cross
			sections for the 4.44 MeV line of $^{12}$C traget, we summed the two components
			at E$_{\gamma}$ = 4.439 MeV and 4.444 MeV (lines of $^{12}$C and
			$^{11}$B, respectively), since it was not possible to separate them
			experimentally. Then, we performed TALYS calculations of $\sigma$(E$_{p}$)
			with our modified OMP and $\beta_{\lambda}$ level deformation parameters
			as input data (in addition to E$_{p}$ and (Z, A) values of the reaction
			partners) for all the ten selected intense lines from Tables (\ref{tab:2},
			\ref{tab:3}). The obtained results for the two main lines of $^{12}$C
			and $^{16}$O, taken as reference lines, are in excellent agreement
			with the experimental data, regarding both the energy dependencies
			and the absolute values of the experimental cross sections. Table
			\ref{tab:7} reports the modified OMP parameters used in our TALYS
			calculations, while we report in Figs. (\ref{fig:10}-\ref{fig:13})
			the $\sigma$(E$_{p}$) excitation curves calculated by TALYS, together
			with the experimental data sets from the current and previous works.
			One can observe, in Figs (\ref{fig:11}, \ref{fig:13}), that the
			theoretical results following the two calculation modes (default or
			modified) for the remaining eight analyzed lines from the two targets,
			at E$_{\gamma}$ = 0.718, 1.022 MeV (of $^{10}$B), 2.000 MeV ($^{11}$C),
			2.124 MeV ($^{11}$B), 2.313 MeV ($^{14}$N), 4.439 MeV ($^{12}$C),
			5.240 MeV ($^{15}$O) and 5.269 MeV ($^{15}$N), agree fairly well,
			in general. The most noticeable differences appear only towards the
			high-energy region for the lines of $^{11}$C and $^{11}$B from the
			$^{nat}$C target and for the line of $^{15}$O from the Mylar target
			at E$_{\gamma}$ = 2.000, 2.124 and 5.240 MeV, respectively. Besides,
			the comparison of TALYS-calculated excitation curves with corresponding
			experimental cross sections shows overall agreements, except for the
			two lines at E$_{\gamma}$ = 1.022 MeV ($^{10}$B line from the C
			target) and E$_{\gamma}$ = 4.439 MeV (line of $^{12}$C from the
			Mylar target via the $^{16}$O(p, p\textquoteright $\alpha$)$^{12}$C
			reaction) that exhibit substantial deviations between the calculated
			and measured absolute $\sigma$-values. Otherwise, while the agreement
			is quite good for the 0.718 MeV line of $^{10}$B from the $^{nat}$C
			target, significant deviations feature the other lines in Figs. (\ref{fig:11},
			\ref{fig:13}) from the two irradiated C and Mylar targets. Table
			\ref{tab:8} lists the observed ratios between TALYS
			code\textendash calculated cross sections and experimental data.

			\subsection{Comparison to the Murphy et al. compilation}
			
			Being of semi-empirical character, the compilation of Murphy, Kozlovsky,
			Kiener and Share \cite{R.J.MurphyB.KozlovskyJ.Kiener2009} is based
			on the available (in 2009) experimental cross section data for main
			$\gamma$-ray lines from low-lying nuclear states, and on TALYS code
			\cite{TALYS} extrapolations to higher projectile energies where experimental
			data were missing. We also report in Figs. (\ref{fig:10}-\ref{fig:13})
			the cross section excitation curves predicted by this database \cite{R.J.MurphyB.KozlovskyJ.Kiener2009}
			for comparison to $\sigma$(E$_{p}$) experimental data from the present
			and previous works.
			
			One expects that regarding the lines for which previous experimental
			cross section data exist, the predicted cross sections should reproduce
			well the measured values. This is indeed true for the main line at
			E$_{\gamma}$ = 4.439 MeV of $^{12}$C produced in (p, p\textquoteright $\gamma$)
			inelastic proton scattering reaction (see Figs \ref{fig:10}), showing
			that the semi-empirical curve from the Murphy et al. compilation \cite{R.J.MurphyB.KozlovskyJ.Kiener2009}
			accounts quite well for nearly all the experimental data points including
			our values up to E$_{p}$ = 200 MeV. Similarly, the semi-empirical
			curve for the 6.129 MeV line of $^{16}$O from the (p, p\textquoteright $\gamma$)
			reaction with this target nucleus seems to account well for most experimental
			data, including for our values at E$_{p}$ = 125 and 150 MeV (see
			Fig. \ref{fig:12}). However, the value of Lang et al. at E$_{p}$
			= 65 MeV and our values at E$_{p}$ = 175 and 200 MeV are not reproduced
			by this curve. For the remaining four lines from the $^{nat}$C target
			(Fig. \ref{fig:11}), the derived semi-empirical curves agree reasonably
			well and better than the excitation curves predicted by TALYS with
			all the experimental data for the lines of $^{10}$B, $^{11}$C and
			$^{11}$B at 0.728, 2.000 and 2.124 MeV, respectively, but not for the
			1.022 MeV line of $^{10}$B. Concerning the other four lines from
			the Mylar targets (see Fig. \ref{fig:13}), one observes that for
			the 4.439 MeV line of $^{12}$C from the $^{16}$O(p, p\textquoteright $\alpha$)$^{12}$C
			reaction, the curve from the Murphy et al. compilation \cite{R.J.MurphyB.KozlovskyJ.Kiener2009}
			reproduces very well the experimental data, including our values at
			E$_{p}$ = 30, 42, 54 and 66 MeV. Then it overestimates our values
			for higher energies of up to E$_{p}$ = 200 MeV. In comparison, the
			TALYS-predicted excitation curve describes our low-energy data up
			to E$_{p}$ = 54 MeV, and deviates significantly from our data (by
			a factor of 2). The excitation curves derived for the 5.240 MeV line
			of $^{15}$O both from the Murphy et al. compilation and via TALYS
			code calculation appear to be roughly consistent with experimental
			data. Besides, one observes that for the 2.313 and 5.269 MeV lines
			of $^{14}$N and $^{15}$N the predicted curves from the Murphy et
			al. compilation underestimate or overestimate, respectively, the experimental
			data beyond our value at E$_{p}$ = 40 MeV, lying substantially farther
			below or above experiment. In contrast, the TALYS-calculated excitation
			curves for these two lines are much closer to experimental data, showing
			better consistency with the latter. More precisely, we provide quantitatively
			in Table \ref{tab:8} the observed differences between the $\gamma$-ray
			excitation functions predicted by the Murphy et al. compilation and
			by nuclear reaction models via TALYS code and experimental calculations
			cross section data.

			\section{Summary, discussion and perspectives\label{sec:Summary,-discussion-and}}
			
			In summary, we report here new experimental $\gamma$-ray production
			cross sections for known main and other lines produced in proton-induced
			nuclear reactions on $^{nat}$C and Mylar targets over the energy
			range, E$_{p}$ = 30 - 200 MeV, using the SSC facility of iThemba
			LABS. These quantities were determined by fitting least squares Legendre
			polynomial expansions to the measured differential cross section angular
			distributions over the explored proton energy range. We also report
			line shape data measured versus E$_{p}$ and $\theta_{lab}$ for the
			line complex observed at E$_{\gamma}$ = 4.44 MeV in irradiations
			of both targets that was analyzed and discussed.
			
			Figs. (\ref{fig:10}, \ref{fig:12}) show that the measured cross sections
			for the prominent, main lines of $^{12}$C and $^{16}$O at 4.439 and
			6.129 MeV are very well reproduced by TALYS code \cite{TALYS} calculations
			and the Murphy et al. compilation \cite{R.J.MurphyB.KozlovskyJ.Kiener2009}
			over the wide proton energy range of E$_{p}$ $\approx$ 6 - 200 MeV.
			This is also the case, to lesser extent, for the lines of $^{10}$B
			and $^{15}$O at E$_{\gamma}$ = 0.718 and 5.240 MeV, respectively.
			In addition, the TALYS-calculated and the Murphy et al. predicted
			excitation functions agree satisfactorily with the experimental data
			over the low-energy region below E$_{p}$ = 30 MeV for all the other
			eight analyzed lines from the two targets, except for the 1.022 MeV
			line of $^{10}$B from the $^{nat}$C target.
			
			This is not surprising since over this energy range, TALYS accounts
			very well for the involved, well formulated and implemented in this
			code, nuclear reaction mechanisms (compound nucleus, direct-reaction,
			pre-equilibrium models), while the semi-empirical compilation of Murphy
			et al. is partly based on those data. In contrast, significant deviations
			were observed between the experimental cross-section data and the
			values predicted by the TALYS calculations and the latter database
			at higher proton energies for several $\gamma$-ray lines (see Figs.
			(\ref{fig:11}, \ref{fig:13}) and Table \ref{tab:8}). As it has
			been confirmed in a recent study \cite{voyles2021proton} of proton-induced
			reactions on Fe, Cu, and Ti from threshold to 55 MeV, state-of-the
			art modern nuclear reaction modeling codes (ALICE, CoH, EMPIRE, TALYS)
			yield results that significantly differ from the cross section experimental
			data, particularly when default input parameters are used for calculating
			their a priori predicted values. At high proton energies, one expects
			the $\gamma$-ray production cross sections to decrease smoothly with
			incident proton energy beyond the compound resonance maximum at E$_{p}$
			$\approx$ 20-30MeV, as can be observed here (see Figs. (\ref{fig:10}
			-\ref{fig:13}) and in our experimental cross section excitation functions
			reported in Ref. \cite{W.Yahia-Cherif2020}). Indeed, the (p, p\textquoteright $\gamma$)
			inelastic scattering cross section comprises a number of different
			reaction mechanisms (compound nucleus, direct, pre-equilibrium, or
			fusion-evaporation reactions) that are strongly energy dependent.
			As the proton energy increases above E$_{p}$ =30-40 MeV, more and
			more nuclear reaction channels involving other reaction mechanisms
			(spallation, and even probably nuclear cascades) become successively
			open. This then makes it legitimate to suspect a still incomplete
			implementation of the reaction mechanisms in modern nuclear reaction
			codes at high energy.
			
			These observations therefore lead one to suggest two ways in order
			to improve the predictions of the nuclear reaction models and obtain
			fair agreements between the measured experimental cross section data
			and the nuclear reaction theoretical models. (i) Using new and previous
			experimental cross section data, and (ii) preferentially using OMP
			level deformation and level density modified parameters in the calculation
			codes. For this purpose, one should extract reliable modified parameters
			from theoretical fits to elastic/inelastic experimental nucleon scattering
			data, as we did for the 4.439 and 6.129 MeV lines of $^{12}$C and
			$^{16}$O. We have also systematically performed such detailed analysis 
			for all the 41 $\gamma$-ray lines from the Mg, Si
			and Fe targets reported in Ref. \cite{W.Yahia-Cherif2020} by means
			of the OPTMAN code \cite{soukhovitskiOPTMAN} for coupled-channels
			nuclear reactions, obtaining fair agreements between TALYS predictions
			and the experimental data. In this respect, the realistic variation
			trends of the experimental $\gamma$-ray production cross sections
			for proton-induced reactions, including our new data sets (see Figs.
			(\ref{fig:10}-\ref{fig:13}) and Ref. \cite{W.Yahia-Cherif2020})
			over the wide proton energy range, E$_{p}$ = 30 - 200 MeV, should
			be usefully taken into account for better implementing these models
			in nuclear calculation codes \cite{M.HermanEMPIRE,W.Yahia-Cherif2020},
			particularly in TALYS. The determination of $\gamma$-ray production
			cross sections for other targets (Mg, Al, Si, Fe) abundant in astrophysical
			sites from our experimental campaigns over the proton energy range
			of E$_{p}$ = 66 \textendash{} 200 MeV, is being achieved. In the
			other hand, new experimental data sets have been recently reported
			by Kiener et al. \cite{Kiener2021PhyRevC} for $\alpha$-particle-induced
			reactions over the incident energy range of E$_{\alpha}$ = 50 \textendash{}
			90 MeV. Those data sets sometimes exhibit significant disagreements
			with TALYS code predictions and with the Murphy et al. compilation. 
			
			This points to the need of performing additional measurements of $\gamma$-ray
			production cross sections in both proton and $\alpha$-particle-induced
			reactions at the SSC facility of iThemba LABS in order to update and
			further enrich the existing data base \cite{R.J.MurphyB.KozlovskyJ.Kiener2009}.
			They could consist in joint collaboration experiments devoted to better
			explore with more observation angles the important 6.129 MeV octupole
			main line of $^{16}$O in proton-induced reactions over the energy
			range, E$_{p}$ = 30 \textendash{} 110 MeV, and for extending the
			recent data sets from Ref. \cite{Kiener2021PhyRevC} to higher $\alpha$-particle
			energies up to the maximum permitted value, E$_{\alpha}$ = 200 MeV.
			To this end, one could take advantage of the recently upgraded AFRODITE
			array equipped with a new frame combining both clover and LaBr$_{3}$:
			Ce $\gamma$-ray detectors \cite{azaiez2020ithemba,msebi2021fast}.
			This improved detection array of high-energy resolution and high efficiency
			would enable one to measure much more complete angular distributions
			for $\gamma$-ray lines of high multipolarity values than this was
			possible with the older detection array used in our experimental campaigns.
			The new $\gamma$-ray line production cross section results reported in this work have various applications including proton therapy and nuclear astrophysics. 
                        Concerning the latter topic, extending the previous low energy data up to E$_{p}$ = 200 MeV, 
                        they will considerably increase the precision in the computed hadron-induced
			$\gamma$-ray flux below E$_{\gamma}$ = 1 GeV, then allowing precision
			modeling \cite{EKafexhiu2018} of solar flare observational data reported
			by the Large Area Telescope onboard the Fermi satellite \cite{atwood2009large}.
			They would also help one elucidating the diffuse $\gamma$-ray emission
			induced in LECRs interactions with the gas and dust of the inner galaxy.
			One could then recalculate with higher accuracy the total nuclear
			$\gamma$-ray flux \cite{benhabiles2013erratum,W.Yahia-Cherif2020}
			from these elusive interactions, the rate of the observed H$_{2}$
			ionization in diffuse interstellar clouds \cite{kiener2012}, and
			maybe constrain the essentially undetermined spectrum of LECRs below
			200 MeV/u \cite{indriolo2009implications}. In this context, the knowledge
			of $\gamma$-ray production cross sections in nuclear reactions from
			threshold up the highest possible energies of the accelerated particles
			delivered by laboratory particle accelerators is of great interest.

			\section*{Acknowledgements}
			
			The authors are indebted to the technical staff of the iThemba LABS
			SSC accelerator for their kind help and friendly cooperation.
			Saad Ouichaoui expresses his thanks and is very grateful to
			Dr R. Nchodu for his warm welcoming of us throughout our experimental campaigns at
			the SSC facility of iThemba LABS. He also warmly thanks Dr F. Azaiez, current Director of 
			iThemba LABS, for his kind hospitality.
			
			This work has been carried out in the framework of a joint scientific cooperation agreement between the USTHB university of Algiers and iThemba LABS of 	
			Cape Town. The work was partially supported by the General Direction of Scientific Research and Technological Development of Algeria (project code A/	
			AS-2013-003), and by the National Research Foundation of South Africa under grants GUN: 109134 and UID87454. Besides, travel support was granted to
 			the collaborating French researchers by the CSNSM and the IPN of Orsay (CNRS/IN2P3 and University of Paris-Sud). Thanks are due to all persons from
 			these institutions who helped in the realisation of this project.

			\bibliographystyle{elsarticle-num}
			\bibliography{ref}

			\section*{Tables}
			
\begin{table}[H]
	{\scriptsize{}}%
	\begin{tabular}{ccccll}
		\hline 
		{\scriptsize{}Target} & {\scriptsize{}Thickness (mg cm$^{-2}$)} & \multicolumn{2}{c}{{\scriptsize{}Areal density, N (atoms cm$^{-2}$)}} & {\scriptsize{}Proton energies $E_{p}$ (MeV); year} & {\scriptsize{}Observation angles (deg); year}\tabularnewline
		\cline{3-4} 
		&  & {\scriptsize{}$^{12}$C} & {\scriptsize{}$^{16}$O} &  & \tabularnewline
		\hline 
		{\scriptsize{}$^{nat}$C} & {\scriptsize{}8.40 $\pm$ 0.07} & {\scriptsize{}0.416 (4)$\times$10$^{21}$} &  & {\scriptsize{}30, 42, 54, 66; 2013.} & {\scriptsize{}85, 95, 130, 140; 2013 + 2015}\tabularnewline
		&  &  &  & {\scriptsize{}66, 80, 95, 110, 125; 2015.} & {\scriptsize{}85, 95, 130, 140, 166, 172; 2016}\tabularnewline
		&  &  &  & {\scriptsize{}66, 125, 150, 175, 200; 2016.} & \tabularnewline
		{\scriptsize{}Mylar (C$_{10}$H$_{8}$O$_{4}$)} & {\scriptsize{}7.00 $\pm$ 0.08} & {\scriptsize{}0.219 (3)$\times$10$^{21}$} & {\scriptsize{}0.088 (1)$\times$10$^{21}$} & {\scriptsize{}30, 42, 54, 66; 2013.} & {\scriptsize{}85, 95, 130, 140; 2013 + 2015}\tabularnewline
		&  &  &  & {\scriptsize{}66, 80, 95, 110, 125; 2015.} & {\scriptsize{}85, 95, 130, 140, 166, 172; 2016}\tabularnewline
		&  &  &  & {\scriptsize{}66, 125, 150, 175, 200; 2016.} & \tabularnewline
		\hline 
	\end{tabular}\caption{Properties of the $^{nat}$C and Mylar targets used in the experiments
	with indication of the observation angles and years. The uncertainties
	on the last digits of their atomic surface density, N, are given in
	parentheses.\label{tab:1}}
\end{table}

		{\footnotesize{}}
		\begin{table}[H]
			\raggedright{}{\footnotesize{}}%
			\begin{tabular}{cccc}
				\hline 
				{\footnotesize{}E$_{\gamma}$(MeV)} & {\footnotesize{}Transition} & {\footnotesize{}Multipolarity} & {\footnotesize{}Reaction}\tabularnewline
				\hline 
				\noalign{\vskip0.3cm}
				{\footnotesize{}0.718} & {\footnotesize{}${{}^1}{{}^0}$B, $1^{+}\thinspace0.718\thinspace\rightarrow\thinspace3^{+}\thinspace g.s$} & {\footnotesize{}E2} & {\footnotesize{}${{}^1}{{}^2}$C(p, X)${{}^1}{{}^0}$B}\tabularnewline
				\noalign{\vskip0.3cm}
				{\footnotesize{}1.022} & {\footnotesize{}${{}^1}{{}^0}$B, $0^{+}\thinspace1.740\thinspace\rightarrow\thinspace1^{+}\thinspace0.718$} & {\footnotesize{}M1} & {\footnotesize{}${{}^1}{{}^2}$C(p, X)${{}^1}{{}^0}$B}\tabularnewline
				\noalign{\vskip0.3cm}
				{\footnotesize{}1.436} & {\footnotesize{}${{}^1}{{}^0}$B, $1^{+}\thinspace2.154\thinspace\rightarrow\thinspace1^{+}\thinspace0.718$} & {\footnotesize{}M1+E2} & {\footnotesize{}${{}^1}{{}^2}$C(p, X)${{}^1}{{}^0}$B}\tabularnewline
				\noalign{\vskip0.3cm}
				{\footnotesize{}2.000} & {\footnotesize{}${{}^1}{{}^1}$C, $\frac{1}{2}^{-}\thinspace2.000\thinspace\rightarrow\thinspace\frac{3}{2}^{-}\thinspace g.s$} & {\footnotesize{}M1} & {\footnotesize{}${{}^1}{{}^2}$C(p, X)${{}^1}{{}^1}$C}\tabularnewline
				\noalign{\vskip0.3cm}
				{\footnotesize{}2.124} & {\footnotesize{}${{}^1}{{}^1}$B, $\frac{1}{2}^{-}\thinspace2.125\thinspace\rightarrow\thinspace\frac{3}{2}^{-}\thinspace g.s$} & {\footnotesize{}M1} & {\footnotesize{}${{}^1}{{}^2}$C(p, 2p\textquoteright )${{}^1}{{}^1}$B}\tabularnewline
				\noalign{\vskip0.3cm}
				{\footnotesize{}2.154} & {\footnotesize{}${{}^1}{{}^0}$B, $1^{+}\thinspace2.154\thinspace\rightarrow\thinspace3^{+}\thinspace g.s$} & {\footnotesize{}E2} & {\footnotesize{}${{}^1}{{}^2}$C(p, X)${{}^1}{{}^0}$B}\tabularnewline
				\noalign{\vskip0.3cm}
				{\footnotesize{}4.319} & {\footnotesize{}${{}^1}{{}^1}$C, $\frac{5}{2}^{-}\thinspace4.319\thinspace\rightarrow\thinspace\frac{3}{2}^{-}\thinspace g.s$} &  & {\footnotesize{}${{}^1}{{}^2}$C(p, X)${{}^1}{{}^1}$C}\tabularnewline
				\noalign{\vskip0.3cm}
				{\footnotesize{}4.439} & {\footnotesize{}${{}^1}{{}^2}$C, $2^{+}\thinspace4.440\thinspace\rightarrow\thinspace0^{+}\thinspace g.s$} & {\footnotesize{}E2} & {\footnotesize{}${{}^1}{{}^2}$C(p,p\textquoteright )${{}^1}{{}^2}$C}\tabularnewline
				\noalign{\vskip0.3cm}
				{\footnotesize{}4.444} & {\footnotesize{}${{}^1}{{}^1}$B, $\frac{5}{2}^{-}\thinspace4.445\thinspace\rightarrow\thinspace\frac{3}{2}^{-}\thinspace g.s$} & {\footnotesize{}M1+E2} & {\footnotesize{}${{}^1}{{}^2}$C(p, 2p\textquoteright )${{}^1}{{}^1}$B}\tabularnewline
				\noalign{\vskip0.3cm}
				{\footnotesize{}6.129} & {\footnotesize{}$^{16}$O, $3^{-}\thinspace6.130\thinspace\rightarrow\thinspace0^{+}\thinspace g.s$} & {\footnotesize{}E3} & {\footnotesize{}${{}^1}{{}^6}$O (p, p\textquoteright ) ${{}^1}{{}^6}$O}\tabularnewline
				\hline 
				\noalign{\vskip0.3cm}
			\end{tabular}{\footnotesize{}\caption{Properties of the $\gamma$-ray transitions observed in proton-induced
				nuclear reactions on the $^{nat}$C target are listed: the $\gamma$-ray
				energy (in the first column), the involved isotopes and the connected
				initial and final states, together with their J$^{\pi}$ values (in
				the second column), the line multipolarity (in the third column),
				and the possible reaction channels leading to the emitting nucleus
				(in column 4). In the latter column, the light reaction products could
				be composed of different combinations of particles.\label{tab:2}}
		}
	\end{table}
	{\footnotesize \par}
	
	{\footnotesize{}}
	\begin{table}[H]
		\raggedright{}{\footnotesize{}}%
		\begin{tabular}{cccc}
			\hline 
			{\footnotesize{}E$_{\gamma}$(MeV)} & {\footnotesize{}Transition} & {\footnotesize{}Multipolarity} & {\footnotesize{}Reaction}\tabularnewline
			\hline 
			\noalign{\vskip0.3cm}
			{\footnotesize{}0.728} & {\footnotesize{}${{}^1}{{}^4}$N, $3^{-}\thinspace5.834\thinspace\rightarrow\thinspace2^{-}\thinspace5.106$} & {\footnotesize{}M1+E2} & {\footnotesize{}${{}^1}{{}^6}$O(p, X)${{}^1}{{}^4}$N}\tabularnewline
			\noalign{\vskip0.3cm}
			{\footnotesize{}1.635} & {\footnotesize{}${{}^1}{{}^4}$N, $1^{+}\thinspace3.948\thinspace\rightarrow\thinspace0^{+}\thinspace2.313$} & {\footnotesize{}M1} & {\footnotesize{}${{}^1}{{}^6}$O(p, X)${{}^1}{{}^4}$N}\tabularnewline
			\noalign{\vskip0.3cm}
			{\footnotesize{}2.313} & {\footnotesize{}${{}^1}{{}^4}$N, $0^{+}\thinspace2.313\thinspace\rightarrow\thinspace1^{+}\thinspace gs$} & {\footnotesize{}M1} & {\footnotesize{}${{}^1}{{}^6}$O(p, X)${{}^1}{{}^4}$N}\tabularnewline
			\noalign{\vskip0.3cm}
			{\footnotesize{}2.498} & {\footnotesize{}${{}^1}{{}^4}$N, $3^{+}\thinspace6.446\thinspace\rightarrow\thinspace1^{+}\thinspace3.948$} & {\footnotesize{}E2} & {\footnotesize{}${{}^1}{{}^6}$O(p, X)${{}^1}{{}^4}$N}\tabularnewline
			\noalign{\vskip0.3cm}
			{\footnotesize{}3.684} & {\footnotesize{}${{}^1}{{}^3}$C, $\frac{3}{2}^{-}\thinspace3.685\thinspace\rightarrow\thinspace\frac{1}{2}^{-}\thinspace g.s$} & {\footnotesize{}M1+E2} & {\footnotesize{}${{}^1}{{}^6}$O(p, X)${{}^1}{{}^3}$C}\tabularnewline
			\noalign{\vskip0.3cm}
			{\footnotesize{}3.853} & {\footnotesize{}${{}^1}{{}^3}$C, $\frac{5}{2}^{+}\thinspace3.854\thinspace\rightarrow\thinspace\frac{1}{2}^{-}\thinspace g.s$} & {\footnotesize{}M2+E3} & {\footnotesize{}${{}^1}{{}^6}$O(p, X)${{}^1}{{}^3}$C}\tabularnewline
			\noalign{\vskip0.3cm}
			{\footnotesize{}4.439} & {\footnotesize{}${{}^1}{{}^2}$C, $2^{+}\thinspace4.440\thinspace\rightarrow\thinspace0^{+}\thinspace g.s$} & {\footnotesize{}E2} & {\footnotesize{}${{}^1}{{}^6}$O(p, X)${{}^1}{{}^2}$C}\tabularnewline
			\noalign{\vskip0.3cm}
			{\footnotesize{}5.105} & {\footnotesize{}${{}^1}{{}^4}$N, $2^{-}\thinspace5.106\thinspace\rightarrow\thinspace1^{+}\thinspace g.s$} & {\footnotesize{}E1+M2+E3} & {\footnotesize{}${{}^1}{{}^6}$O(p, X)${{}^1}{{}^4}$N}\tabularnewline
			\noalign{\vskip0.3cm}
			{\footnotesize{}5.240} & {\footnotesize{}${{}^1}{{}^5}$O, $\frac{5}{2}^{+}\thinspace5.241\thinspace\rightarrow\thinspace\frac{1}{2}^{-}\thinspace g.s$} & {\footnotesize{}M2+E3} & {\footnotesize{}${{}^1}{{}^6}$O(p, X)${{}^1}{{}^5}$O}\tabularnewline
			\noalign{\vskip0.3cm}
			{\footnotesize{}5.269} & {\footnotesize{}${{}^1}{{}^5}$N, $\frac{5}{2}^{+}\thinspace5.270\thinspace\rightarrow\thinspace\frac{1}{2}^{-}\thinspace g.s$} & {\footnotesize{}M2+E3} & {\footnotesize{}${{}^1}{{}^6}$O(p, 2p\textquoteright )${{}^1}{{}^5}$N}\tabularnewline
			\noalign{\vskip0.3cm}
			{\footnotesize{}6.129} & {\footnotesize{}$^{16}$O, $3^{-}\thinspace6.130\thinspace\rightarrow\thinspace0^{+}\thinspace g.s$} & {\footnotesize{}E3} & {\footnotesize{}${{}^1}{{}^6}$O (p, p\textquoteright ) ${{}^1}{{}^6}$O}\tabularnewline
			\hline 
			\noalign{\vskip0.3cm}
		\end{tabular}{\footnotesize{}\caption{Same as in TABLE \ref{tab:2} but for the Mylar target. All the reactions
			in column 4 are induced on the $^{16}$O nucleus.\label{tab:3}}
	}
\end{table}
{\footnotesize \par}

{\footnotesize{}}
\begin{table}[H]
	\raggedright{}{\footnotesize{}}%
	\begin{tabular}{cc}
		\hline 
		{\footnotesize{}E$_{\gamma}$(MeV)} & {\footnotesize{}Origine }\tabularnewline
		\hline 
		{\footnotesize{}0.198} & {\footnotesize{}$^{70}$Ge(n, $\gamma$)$^{71}$Ge}\tabularnewline
		{\footnotesize{}0.511} & {\footnotesize{}$e^{+}e^{-}$annihilation}\tabularnewline
		{\footnotesize{}0.596} & {\footnotesize{}$^{74}$Ge(n,n\textquoteright{} $\gamma$)$^{74}$Ge}\tabularnewline
		{\footnotesize{}0.689} & {\footnotesize{}$^{72}$Ge(n,n\textquoteright{})$^{72}$Ge}\tabularnewline
		{\footnotesize{}0.834} & {\footnotesize{}$^{72}$Ge(n,n\textquoteright{} $\gamma$)$^{72}$Ge}\tabularnewline
		{\footnotesize{}0.844} & {\footnotesize{}$^{27}$Al(p,p\textquoteright{} $\gamma$)$^{27}$Al}\tabularnewline
		{\footnotesize{}0.868} & {\footnotesize{}$^{74}$Ge(n,n\textquoteright{} $\gamma$)$^{74}$Ge}\tabularnewline
		{\footnotesize{}0.894} & {\footnotesize{}$^{72}$Ge(n,n\textquoteright{} $\gamma$)$^{72}$Ge}\tabularnewline
		{\footnotesize{}1.014} & {\footnotesize{}$^{27}$Al(p,p\textquoteright{} $\gamma$)$^{27}$Al}\tabularnewline
		{\footnotesize{}1.369} & {\footnotesize{}$^{27}$Al(p, X)$^{24}$Mg}\tabularnewline
		{\footnotesize{}1.612} & {\footnotesize{}$^{27}$Al(p, X)$^{25}$Mg}\tabularnewline
		{\footnotesize{}1.809} & {\footnotesize{}$^{27}$Al(p,2p\textquoteright{} $\gamma$)$^{26}$Mg}\tabularnewline
		{\footnotesize{}2.212} & {\footnotesize{}$^{27}$Al(p,p\textquoteright{} $\gamma$)$^{27}$Al}\tabularnewline
		{\footnotesize{}2.754} & {\footnotesize{}$^{27}$Al(p,$\alpha$ $\gamma$)$^{24}$Mg}\tabularnewline
		{\footnotesize{}3.004} & {\footnotesize{}$^{27}$Al(p,p\textquoteright{} $\gamma$)$^{27}$Al}\tabularnewline
		\hline 
	\end{tabular}{\footnotesize{}\caption{Background $\gamma$-ray lines produced in proton beam interactions
		with surrounding materials (Al, Ge) forming the reaction chamber and
		HP-Ge detectors\label{tab:4}}
}
\end{table}
{\footnotesize \par}

{\footnotesize{}}
\begin{table}[H]
	\textit{\tiny{}}%
	\begin{tabular}{ccccccccccccccccc}
		\hline 
		\multicolumn{17}{c}{\textit{\tiny{}p + $^{nat}$C}}\tabularnewline
		& \multicolumn{3}{c}{\textit{\tiny{}0.718 MeV}} & \multicolumn{3}{l}{\textit{\tiny{}1.022 MeV}} & \multicolumn{3}{l}{\textit{\tiny{}2.00 MeV}} & \multicolumn{3}{l}{\textit{\tiny{}2.124 MeV}} & \multicolumn{4}{c}{\textit{\tiny{}4.44 MeV}}\tabularnewline
		\textit{\tiny{}E$_{p}$(MeV)} & \textit{\tiny{}4$\pi$a$_{0}$} & \textit{\tiny{}4$\pi$a$_{2}$} & \textit{\tiny{}4$\pi$a$_{4}$} & \textit{\tiny{}4$\pi$a$_{0}$} & \textit{\tiny{}4$\pi$a$_{2}$} &  & \textit{\tiny{}4$\pi$a$_{0}$} & \textit{\tiny{}4$\pi$a$_{2}$} &  & \textit{\tiny{}4$\pi$a$_{0}$} & \textit{\tiny{}4$\pi$a$_{2}$} &  & \textit{\tiny{}4$\pi$a$_{0}$} & \textit{\tiny{}4$\pi$a$_{2}$} & \textit{\tiny{}4$\pi$a$_{4}$} & \tabularnewline
		\hline 
		\textit{\tiny{}30} & \textit{\tiny{}7.81} & \textit{\tiny{}-2.38} & \textit{\tiny{}-1.89} & \textit{\tiny{}1.27} & \textit{\tiny{}-0.55} &  & \textit{\tiny{}19.49} & \textit{\tiny{}-1.93} &  & \textit{\tiny{}8.47} & \textit{\tiny{}-1.57} &  & \textit{\tiny{}71.18} & \textit{\tiny{}-6.44} & \textit{\tiny{}-32.64} & \tabularnewline
		\textit{\tiny{}42} & \textit{\tiny{}14.76} & \textit{\tiny{}-4.33} & \textit{\tiny{}-3.62} & \textit{\tiny{}1.55} & \textit{\tiny{}-0.57} &  & \textit{\tiny{}14.80} & \textit{\tiny{}2.10} &  & \textit{\tiny{}6.52} & \textit{\tiny{}1.40} &  & \textit{\tiny{}53.87} & \textit{\tiny{}-21.75} & \textit{\tiny{}-27.23} & \tabularnewline
		\textit{\tiny{}54} & \textit{\tiny{}13.58} & \textit{\tiny{}-4.06} & \textit{\tiny{}-4.26} & \textit{\tiny{}1.19} & \textit{\tiny{}-0.56} &  & \textit{\tiny{}9.66} & \textit{\tiny{}-0.02} &  & \textit{\tiny{}3.56} & \textit{\tiny{}0.29} &  & \textit{\tiny{}25.57} & \textit{\tiny{}-7.07} & \textit{\tiny{}-2.54} & \tabularnewline
		\textit{\tiny{}66} & \textit{\tiny{}8.13} & \textit{\tiny{}-0.80} & \textit{\tiny{}1.76} & \textit{\tiny{}1.12} & \textit{\tiny{}-0.42} &  & \textit{\tiny{}7.34} & \textit{\tiny{}0.09} &  & \textit{\tiny{}2.65} & \textit{\tiny{}0.06} &  & \textit{\tiny{}17.60} & \textit{\tiny{}-3.00} & \textit{\tiny{}0.37 } & \tabularnewline
		\textit{\tiny{}80} & \textit{\tiny{}9.05} & \textit{\tiny{}-0.90} & \textit{\tiny{}-1.07} & \textit{\tiny{}0.96} & \textit{\tiny{}-0.01} &  & \textit{\tiny{}6.34} & \textit{\tiny{}-0.24} &  & \textit{\tiny{}2.67} & \textit{\tiny{}-0.13} &  & \textit{\tiny{}13.09} & \textit{\tiny{}-3.62} & \textit{\tiny{}0.64} & \tabularnewline
		\textit{\tiny{}95} & \textit{\tiny{}8.45} & \textit{\tiny{}-0.53} & \textit{\tiny{}-0.76} & \textit{\tiny{}0.88} & \textit{\tiny{}-0.12} &  & \textit{\tiny{}5.67} & \textit{\tiny{}0.52} &  & \textit{\tiny{}2.34} & \textit{\tiny{}0.67} &  & \textit{\tiny{}12.55} & \textit{\tiny{}-0.87} & \textit{\tiny{}2.88 } & \tabularnewline
		\textit{\tiny{}110} & \textit{\tiny{}8.25} & \textit{\tiny{}-0.57} & \textit{\tiny{}-1.19} & \textit{\tiny{}0.88} & \textit{\tiny{}-0.03} &  & \textit{\tiny{}5.25} & \textit{\tiny{}0.66} &  & \textit{\tiny{}2.36} & \textit{\tiny{}0.66} &  & \textit{\tiny{}9.98} & \textit{\tiny{}-1.64} & \textit{\tiny{}2.93} & \tabularnewline
		\textit{\tiny{}125} & \textit{\tiny{}7.14} & \textit{\tiny{}0.00} & \textit{\tiny{}0.84} & \textit{\tiny{}0.79} & \textit{\tiny{}-0.16} &  & \textit{\tiny{}5.31} & \textit{\tiny{}0.21} &  & \textit{\tiny{}2.28} & \textit{\tiny{}0.37} &  & \textit{\tiny{}9.78} & \textit{\tiny{}-3.79} & \textit{\tiny{}-0.46 } & \tabularnewline
		\textit{\tiny{}150} & \textit{\tiny{}7.41} & \textit{\tiny{}-1.41} & \textit{\tiny{}-0.12} & \textit{\tiny{}0.77} & \textit{\tiny{}-0.03} &  & \textit{\tiny{}5.51} & \textit{\tiny{}1.02} &  & \textit{\tiny{}2.63} & \textit{\tiny{}0.68} &  & \textit{\tiny{}8.13} & \textit{\tiny{}-3.55} & \textit{\tiny{}1.58 } & \tabularnewline
		\textit{\tiny{}175} & \textit{\tiny{}6.96} & \textit{\tiny{}-1.01} & \textit{\tiny{}-1.10} & \textit{\tiny{}0.74} & \textit{\tiny{}-0.03} &  & \textit{\tiny{}4.71} & \textit{\tiny{}0.50} &  & \textit{\tiny{}2.24} & \textit{\tiny{}0.42} &  & \textit{\tiny{}7.59} & \textit{\tiny{}-3.81} & \textit{\tiny{}0.34 } & \tabularnewline
		\textit{\tiny{}200} & \textit{\tiny{}6.69} & \textit{\tiny{}-0.40} & \textit{\tiny{}-1.25} & \textit{\tiny{}0.72} & \textit{\tiny{}-0.01} &  & \textit{\tiny{}3.76} & \textit{\tiny{}0.40} &  & \textit{\tiny{}1.86} & \textit{\tiny{}0.51} &  & \textit{\tiny{}6.48} & \textit{\tiny{}-2.68} & \textit{\tiny{}0.96 } & \tabularnewline
		\multicolumn{17}{c}{\textit{\tiny{}p + Mylar}}\tabularnewline
		& \multicolumn{3}{l}{\textit{\tiny{}2.313 MeV}} & \multicolumn{3}{c}{\textit{\tiny{}4.439 MeV}} & \multicolumn{3}{c}{\textit{\tiny{}5.240 MeV}} & \multicolumn{3}{c}{\textit{\tiny{}5.270 MeV}} & \multicolumn{4}{c}{\textit{\tiny{}6.129 MeV}}\tabularnewline
		\textit{\tiny{}E$_{p}$(MeV)} & \textit{\tiny{}4$\pi$a$_{0}$} & \textit{\tiny{}4$\pi$a$_{2}$} &  & \textit{\tiny{}4$\pi$a$_{0}$} & \textit{\tiny{}4$\pi$a$_{2}$} & \textit{\tiny{}4$\pi$a$_{4}$} & \textit{\tiny{}4$\pi$a$_{0}$} & \textit{\tiny{}4$\pi$a$_{2}$} & \textit{\tiny{}4$\pi$a$_{4}$} & \textit{\tiny{}4$\pi$a$_{0}$} & \textit{\tiny{}4$\pi$a$_{2}$} & \textit{\tiny{}4$\pi$a$_{4}$} & \textit{\tiny{}4$\pi$a$_{0}$} & \textit{\tiny{}4$\pi$a$_{2}$} & \textit{\tiny{}4$\pi$a$_{4}$} & \textit{\tiny{}4$\pi$a$_{6}$}\tabularnewline
		\hline 
		\textit{\tiny{}30} & \textit{\tiny{}16.74 } & \textit{\tiny{}0.93} &  & \textit{\tiny{}93.55} & \textit{\tiny{}26.02} & \textit{\tiny{}20.84} & \textit{\tiny{}13.18} & \textit{\tiny{}-3.05} & \textit{\tiny{}-5.17} & \textit{\tiny{}10.31} & \textit{\tiny{}-3.92} & \textit{\tiny{}-2.91} &  &  &  & \tabularnewline
		\textit{\tiny{}42} & \textit{\tiny{}17.87} & \textit{\tiny{}-2.24} &  & \textit{\tiny{}48.23} & \textit{\tiny{}28.10} & \textit{\tiny{}23.03} & \textit{\tiny{}6.83} & \textit{\tiny{}-1.95} & \textit{\tiny{}-3.06} & \textit{\tiny{}4.44} & \textit{\tiny{}-1.76} & \textit{\tiny{}-2.18} &  &  &  & \tabularnewline
		\textit{\tiny{}54} & \textit{\tiny{}20.42} & \textit{\tiny{}-3.11} &  & \textit{\tiny{}29.76} & \textit{\tiny{}2.22} & \textit{\tiny{}5.89} & \textit{\tiny{}6.27} & \textit{\tiny{}-2.14} & \textit{\tiny{}-2.56} & \textit{\tiny{}3.84} & \textit{\tiny{}-1.16} & \textit{\tiny{}-1.03} &  &  &  & \tabularnewline
		\textit{\tiny{}66} & \textit{\tiny{}14.07} & \textit{\tiny{}-2.50} &  & \textit{\tiny{}18.07} & \textit{\tiny{}-5.98} & \textit{\tiny{}-5.67 } & \textit{\tiny{}6.21} & \textit{\tiny{}2.18} & \textit{\tiny{}0.28} & \textit{\tiny{}3.66} & \textit{\tiny{}0.95} & \textit{\tiny{}-0.47} & \textit{\tiny{}13.53} & \textit{\tiny{}-0.63} & \textit{\tiny{}-0.12} & \textit{\tiny{}0.69}\tabularnewline
		\textit{\tiny{}80} & \textit{\tiny{}12.06} & \textit{\tiny{}-1.71} &  & \textit{\tiny{}8.56} & \textit{\tiny{}0.36} & \textit{\tiny{}5.32} & \textit{\tiny{}4.12} & \textit{\tiny{}-2.13} & \textit{\tiny{}-1.86} & \textit{\tiny{}2.45} & \textit{\tiny{}-0.38} & \textit{\tiny{}-0.44 } &  &  &  & \tabularnewline
		\textit{\tiny{}95} & \textit{\tiny{}11.70} & \textit{\tiny{}-0.60} &  & \textit{\tiny{}4.12} & \textit{\tiny{}-2.07} & \textit{\tiny{}-2.69} & \textit{\tiny{}3.67} & \textit{\tiny{}0.43} & \textit{\tiny{}0.43} & \textit{\tiny{}2.05} & \textit{\tiny{}-0.20} & \textit{\tiny{}-0.08} &  &  &  & \tabularnewline
		\textit{\tiny{}110} & \textit{\tiny{}10.52} & \textit{\tiny{}-1.43} &  & \textit{\tiny{}3.91} & \textit{\tiny{}-1.36} & \textit{\tiny{}-0.32} & \textit{\tiny{}3.11} & \textit{\tiny{}0.01} & \textit{\tiny{}0.03} & \textit{\tiny{}1.86} & \textit{\tiny{}0.13} & \textit{\tiny{}0.15} &  &  &  & \tabularnewline
		\textit{\tiny{}125} & \textit{\tiny{}10.41} & \textit{\tiny{}0.82} &  & \textit{\tiny{}3.82} & \textit{\tiny{}1.47} & \textit{\tiny{}1.24} & \textit{\tiny{}3.08} & \textit{\tiny{}-0.76} & \textit{\tiny{}-0.34} & \textit{\tiny{}1.88} & \textit{\tiny{}-1.77} & \textit{\tiny{}-1.55} & \textit{\tiny{}5.40} & \textit{\tiny{}-1.98} & \textit{\tiny{}0.42} & \textit{\tiny{}1.85}\tabularnewline
		\textit{\tiny{}150} & \textit{\tiny{}10.82} & \textit{\tiny{}0.56} &  & \textit{\tiny{}3.58} & \textit{\tiny{}0.70} & \textit{\tiny{}1.89} & \textit{\tiny{}3.16} & \textit{\tiny{}-0.12} & \textit{\tiny{}-0.23} & \textit{\tiny{}1.93} & \textit{\tiny{}-0.12} & \textit{\tiny{}-0.35} & \textit{\tiny{}4.90} & \textit{\tiny{}-1.30} & \textit{\tiny{}0.11} & \textit{\tiny{}0.91}\tabularnewline
		\textit{\tiny{}175} & \textit{\tiny{}9.24} & \textit{\tiny{}-0.78} &  & \textit{\tiny{}2.58} & \textit{\tiny{}1.54} & \textit{\tiny{}-1.01} & \textit{\tiny{}2.39} & \textit{\tiny{}-0.10} & \textit{\tiny{}-0.28} & \textit{\tiny{}1.47} & \textit{\tiny{}-0.76} & \textit{\tiny{}0.56} & \textit{\tiny{}4.33} & \textit{\tiny{}-1.59} & \textit{\tiny{}0.81} & \textit{\tiny{}2.05}\tabularnewline
		\textit{\tiny{}200} & \textit{\tiny{}9.07} & \textit{\tiny{}-0.12} &  & \textit{\tiny{}2.49} & \textit{\tiny{}1.37} & \textit{\tiny{}0.18} & \textit{\tiny{}2.92} & \textit{\tiny{}-0.83} & \textit{\tiny{}-0.96} & \textit{\tiny{}1.67} & \textit{\tiny{}-0.09} & \textit{\tiny{}-0.02} & \textit{\tiny{}3.86} & \textit{\tiny{}-0.13} & \textit{\tiny{}0.05} & \textit{\tiny{}0.00}\tabularnewline
		\hline 
	\end{tabular}{\footnotesize{}\caption{List of a$_{\ell}$ coefficients obtained from Legendre polynomial
		expansion (Eq. \ref{eq:2}) fits to the experimental $\gamma$-ray
		angular distributions.\label{tab:5}}
}
\end{table}

\begin{table}[H]
	{\scriptsize{}}%
	\begin{tabular}{ccccccccccccc}
		\hline 
		& \multicolumn{11}{c}{{\scriptsize{}$\sigma(mb)$}} & \tabularnewline
		& \multicolumn{5}{c}{{\scriptsize{}p + $^{nat}$C, E$_{\gamma}$ (MeV)}} &  & \multicolumn{5}{c}{{\scriptsize{}p + Mylar, E$_{\gamma}$ (MeV)}} & \tabularnewline
		\cline{2-6} \cline{8-12} 
		{\scriptsize{}E$_{p}$(MeV)} & {\scriptsize{}0.718} & {\scriptsize{}1.022} & {\scriptsize{}2.00} & {\scriptsize{}2.124} & {\scriptsize{}4.44} &  & {\scriptsize{}2.313} & {\scriptsize{}4.44} & {\scriptsize{}5.240} & {\scriptsize{}5.270} & {\scriptsize{}6.129} & \tabularnewline
		\hline 
		{\scriptsize{}30} & {\scriptsize{}7.8 $\pm$1.4} & {\scriptsize{}1.27 $\pm$0.24} & \textit{\scriptsize{}19.5 }{\scriptsize{}$\pm$4.6} & \textit{\scriptsize{}8.5 }{\scriptsize{}$\pm$1.8} & \textit{\scriptsize{}71.1 }{\scriptsize{}$\pm$15.0} &  & \textit{\scriptsize{}16.7 }{\scriptsize{}$\pm$3.4} & \textit{\scriptsize{}93.5 }{\scriptsize{}$\pm$27.0 } & \textit{\scriptsize{}13.2 }{\scriptsize{}$\pm$2.5} & \textit{\scriptsize{}10.3 }{\scriptsize{}$\pm$2.9} &  & \tabularnewline
		{\scriptsize{}42} & {\scriptsize{}14.8 $\pm$2.5} & {\scriptsize{}1.55 $\pm$0.31} & \textit{\scriptsize{}14.8 }{\scriptsize{}$\pm$3.2} & \textit{\scriptsize{}6.5 }{\scriptsize{}$\pm$1.3} & \textit{\scriptsize{}53.9 }{\scriptsize{}$\pm$12.0} &  & \textit{\scriptsize{}17.9 }{\scriptsize{}$\pm$3.6} & \textit{\scriptsize{}48.2 }{\scriptsize{}$\pm$12.0} & \textit{\scriptsize{}6.8 }{\scriptsize{}$\pm$1.3} & \textit{\scriptsize{}4.4 }{\scriptsize{}$\pm$0.9} &  & \tabularnewline
		{\scriptsize{}54} & {\scriptsize{}13.6 $\pm$2.4} & {\scriptsize{}1.19 $\pm$0.21} & \textit{\scriptsize{}9.7 }{\scriptsize{}$\pm$2.5} & \textit{\scriptsize{}3.6 }{\scriptsize{}$\pm$0.8} & \textit{\scriptsize{}25.6 }{\scriptsize{}$\pm$7.0} &  & \textit{\scriptsize{}20.4 }{\scriptsize{}$\pm$4.2} & \textit{\scriptsize{}29.8 }{\scriptsize{}$\pm$9.5} & \textit{\scriptsize{}6.3 }{\scriptsize{}$\pm$1.2} & \textit{\scriptsize{}3.8 }{\scriptsize{}$\pm$0.8} &  & \tabularnewline
		{\scriptsize{}66} & {\scriptsize{}8.1 $\pm$1.5} & {\scriptsize{}1.12 $\pm$0.20} & \textit{\scriptsize{}7.3 }{\scriptsize{}$\pm$1.9} & \textit{\scriptsize{}2.6 }{\scriptsize{}$\pm$0.6} & \textit{\scriptsize{}17.6 }{\scriptsize{}$\pm$4.3} &  & \textit{\scriptsize{}14.0 }{\scriptsize{}$\pm$2.9} & \textit{\scriptsize{}18.0 }{\scriptsize{}$\pm$5.9} & \textit{\scriptsize{}6.2 }{\scriptsize{}$\pm$1.2} & \textit{\scriptsize{}3.6 }{\scriptsize{}$\pm$0.7} & \textit{\scriptsize{}13.5 }{\scriptsize{}$\pm$2.5} & \tabularnewline
		{\scriptsize{}80} & {\scriptsize{}9.0 $\pm$1.9} & {\scriptsize{}0.96 $\pm$0.18} & \textit{\scriptsize{}6.3 }{\scriptsize{}$\pm$1.5} & \textit{\scriptsize{}2.6 }{\scriptsize{}$\pm$0.7} & \textit{\scriptsize{}13.0 }{\scriptsize{}$\pm$3.7} &  & \textit{\scriptsize{}12.0 }{\scriptsize{}$\pm$2.7} & \textit{\scriptsize{}8.5 }{\scriptsize{}$\pm$2.7} & \textit{\scriptsize{}4.1 }{\scriptsize{}$\pm$0.9} & \textit{\scriptsize{}2.4 }{\scriptsize{}$\pm$0.5} &  & \tabularnewline
		{\scriptsize{}95} & {\scriptsize{}8.4 $\pm$1.6} & {\scriptsize{}0.88 $\pm$0.19} & \textit{\scriptsize{}5.7 }{\scriptsize{}$\pm$1.3} & \textit{\scriptsize{}2.3 }{\scriptsize{}$\pm$0.6} & \textit{\scriptsize{}12.5 }{\scriptsize{}$\pm$3} &  & \textit{\scriptsize{}11.7 }{\scriptsize{}$\pm$2.3} & \textit{\scriptsize{}4.1 }{\scriptsize{}$\pm$1.2} & \textit{\scriptsize{}3.7 }{\scriptsize{}$\pm$0.7} & \textit{\scriptsize{}2.0 }{\scriptsize{}$\pm$0.4} &  & \tabularnewline
		{\scriptsize{}110} & {\scriptsize{}8.2 $\pm$1.4} & {\scriptsize{}0.88 $\pm$0.18} & \textit{\scriptsize{}5.3 }{\scriptsize{}$\pm$1.2} & \textit{\scriptsize{}2.3 }{\scriptsize{}$\pm$0.6} & \textit{\scriptsize{}10.0 }{\scriptsize{}$\pm$2} &  & \textit{\scriptsize{}10.5 }{\scriptsize{}$\pm$1.9} & \textit{\scriptsize{}3.9 }{\scriptsize{}$\pm$1.1} & \textit{\scriptsize{}3.1 }{\scriptsize{}$\pm$0.8} & \textit{\scriptsize{}1.8 }{\scriptsize{}$\pm$0.4} &  & \tabularnewline
		{\scriptsize{}125} & {\scriptsize{}7.1 $\pm$1.4} & {\scriptsize{}0.79 $\pm$0.17} & \textit{\scriptsize{}5.3 }{\scriptsize{}$\pm$1.5} & \textit{\scriptsize{}2.2 }{\scriptsize{}$\pm$0.6} & \textit{\scriptsize{}9.8 }{\scriptsize{}$\pm$1.9} &  & \textit{\scriptsize{}10.4 }{\scriptsize{}$\pm$1.9} & \textit{\scriptsize{}3.8 }{\scriptsize{}$\pm$1.1} & \textit{\scriptsize{}3.0 }{\scriptsize{}$\pm$0.6} & \textit{\scriptsize{}1.8 }{\scriptsize{}$\pm$0.4} & \textit{\scriptsize{}5.4 }{\scriptsize{}$\pm$1.2} & \tabularnewline
		{\scriptsize{}150} & {\scriptsize{}7.4 $\pm$1.7} & {\scriptsize{}0.77 $\pm$0.17} & \textit{\scriptsize{}5.5 }{\scriptsize{}$\pm$1.5} & \textit{\scriptsize{}2.6 }{\scriptsize{}$\pm$0.7} & \textit{\scriptsize{}8.1 }{\scriptsize{}$\pm$2.2} &  & \textit{\scriptsize{}10.8 }{\scriptsize{}$\pm$2.2} & \textit{\scriptsize{}3.5 }{\scriptsize{}$\pm$1.0} & \textit{\scriptsize{}3.1 }{\scriptsize{}$\pm$0.7} & \textit{\scriptsize{}1.9 }{\scriptsize{}$\pm$0.5} & \textit{\scriptsize{}4.90 }{\scriptsize{}$\pm$1.0} & \tabularnewline
		{\scriptsize{}175} & {\scriptsize{}6.9 $\pm$1.2} & {\scriptsize{}0.74 $\pm$0.17} & \textit{\scriptsize{}4.7 }{\scriptsize{}$\pm$1.2} & \textit{\scriptsize{}2.2 }{\scriptsize{}$\pm$0.6} & \textit{\scriptsize{}7.6 }{\scriptsize{}$\pm$1.8} &  & \textit{\scriptsize{}9.2 }{\scriptsize{}$\pm$1.9} & \textit{\scriptsize{}2.5 }{\scriptsize{}$\pm$0.8} & \textit{\scriptsize{}2.4 }{\scriptsize{}$\pm$0.5} & \textit{\scriptsize{}1.4 }{\scriptsize{}$\pm$0.3} & \textit{\scriptsize{}4.3 }{\scriptsize{}$\pm$0.9} & \tabularnewline
		{\scriptsize{}200} & {\scriptsize{}6.6 $\pm$1.2} & {\scriptsize{}0.72 $\pm$0.17} & \textit{\scriptsize{}3.7 }{\scriptsize{}$\pm$1.0} & \textit{\scriptsize{}1.8 }{\scriptsize{}$\pm$0.5} & \textit{\scriptsize{}6.4 }{\scriptsize{}$\pm$1.7} &  & \textit{\scriptsize{}9.0 }{\scriptsize{}$\pm$1.8} & \textit{\scriptsize{}2.4 }{\scriptsize{}$\pm$0.7} & \textit{\scriptsize{}2.5 }{\scriptsize{}$\pm$0.6} & \textit{\scriptsize{}1.6 }{\scriptsize{}$\pm$0.4} & \textit{\scriptsize{}3.8 }{\scriptsize{}$\pm$0.8} & \tabularnewline
		\hline 
	\end{tabular}{\scriptsize \par}
	
	\caption{Experimental values and related uncertainties of integrated cross sections for the identified
		ten most intense $\gamma$-ray lines emitted in proton irradiations
		of the $^{nat}$C and Mylar targets (see the text in Subsection \ref{sub:-ray-line-production} for more details).\label{tab:6}}
\end{table}

{\footnotesize \par}

\begin{table}[H]
	{\small{}}%
	\begin{tabular}{ccc}
		\hline 
		{\small{}Parameters} & {\small{}$^{12}$C} & {\small{}$^{16}$O}\tabularnewline
		\hline 
		{\small{}E$_{f}$ (MeV)} & {\small{}-8.95} & {\small{}-6.36}\tabularnewline
		{\small{}r$_{v}$ (fm)} & {\small{}1.162} & {\small{}1.162}\tabularnewline
		{\small{}a$_{v}$ (fm)} & {\small{}0.665} & {\small{}0.665}\tabularnewline
		{\small{}V$_{1}$ (MeV)} & {\small{}55.4} & {\small{}99.8}\tabularnewline
		{\small{}V$_{2}$ (MeV$^{-1}$)} & {\small{}0.007} & {\small{}0.007}\tabularnewline
		{\small{}V$_{3}$ (MeV$^{-2}$)} & {\small{}0.000027} & {\small{}0.000022}\tabularnewline
		{\small{}W$_{1}$ (MeV)} & {\small{}15.20} & {\small{}15.20}\tabularnewline
		{\small{}W$_{2}$ (MeV)} & {\small{}85.0} & {\small{}75}\tabularnewline
		{\small{}r$_{vd}$ (fm)} & {\small{}1.29} & {\small{}1.29}\tabularnewline
		{\small{}a$_{vd}$ (fm)} & {\small{}0.51} & {\small{}0.51}\tabularnewline
		{\small{}d$_{1}$ (MeV)} & {\small{}14.60} & {\small{}14.60}\tabularnewline
		{\small{}d$_{2}$ (MeV$^{-1}$)} & {\small{}0.0224} & {\small{}0.0224}\tabularnewline
		{\small{}d$_{3}$ (MeV)} & {\small{}11.50} & {\small{}11.50}\tabularnewline
		{\small{}r$_{vso}$ (fm)} & {\small{}1.0} & {\small{}1}\tabularnewline
		{\small{}a$_{vso}$ (fm)} & {\small{}0.58} & {\small{}0.58}\tabularnewline
		{\small{}V$_{so1}$(MeV)} & {\small{}9.50} & {\small{}6.0}\tabularnewline
		{\small{}V$_{so2}$ (MeV$^{-1}$)} & {\small{}0.0136} & {\small{}0.0035}\tabularnewline
		{\small{}W$_{so1}$ (MeV)} & {\small{}-3.1} & {\small{}-3.1}\tabularnewline
		{\small{}W$_{so2}$ (MeV)} & {\small{}160.0} & {\small{}160}\tabularnewline
		\hline 
	\end{tabular}\caption{Adjusted values of the OMP parameters used as modified input data
	in TALYS code calculations of $\gamma$-ray production cross sections
	for proton reactions with $^{12}$C and $^{16}$O target nuclei.\label{tab:7}}
\end{table}

{\footnotesize{}}
\begin{table}[H]
	\raggedright{}{\footnotesize{}}%
	\begin{tabular}{cccccc}
		\hline 
		\multirow{2}{*}{{\footnotesize{}E$_{\gamma}$(MeV)}} & \multicolumn{2}{c}{{\footnotesize{}Talys/Experiment}} &  & \multicolumn{2}{c}{{\footnotesize{}Murphy }\textit{\footnotesize{}et al.}{\footnotesize{}
				/Experiment}}\tabularnewline
		\cline{2-6} 
		\noalign{\vskip0.2cm}
		& {\footnotesize{}E$_{p}$= 66 MeV} & {\footnotesize{}E$_{p}$= 150 MeV} &  & {\footnotesize{}E$_{p}$= 66 MeV} & {\footnotesize{}E$_{p}$= 150 MeV}\tabularnewline
		\hline 
		\noalign{\vskip0.2cm}
		&  &  & {\footnotesize{}p + $^{nat}$C} &  & \tabularnewline
		\noalign{\vskip0.2cm}
		{\footnotesize{}0.718} & {\footnotesize{}2.41 } & {\footnotesize{}0.98} &  & {\footnotesize{}1.96} & {\footnotesize{}1.07}\tabularnewline
		\noalign{\vskip0.2cm}
		{\footnotesize{}1.022} & {\footnotesize{}4.83 } & {\footnotesize{}2.50} &  & {\footnotesize{}5.71} & {\footnotesize{}4.15}\tabularnewline
		\noalign{\vskip0.2cm}
		{\footnotesize{}2.00} & {\footnotesize{}1.05 } & {\footnotesize{}0.37} &  & {\footnotesize{}1.34} & {\footnotesize{}1.21}\tabularnewline
		\noalign{\vskip0.2cm}
		{\footnotesize{}2.124} & {\footnotesize{}1.78 } & {\footnotesize{}0.51} &  & {\footnotesize{}1.74} & {\footnotesize{}1.25}\tabularnewline
		\noalign{\vskip0.2cm}
		{\footnotesize{}4.44} & {\footnotesize{}1.11 } & {\footnotesize{}1.05} &  & {\footnotesize{}1.07} & {\footnotesize{}1.10}\tabularnewline
		\noalign{\vskip0.2cm}
		&  &  & {\footnotesize{}p + Mylar} &  & \tabularnewline
		\noalign{\vskip0.2cm}
		{\footnotesize{}2.313} & {\footnotesize{}0.75 } & {\footnotesize{}0.41} &  & {\footnotesize{}0.40} & {\footnotesize{}0.18}\tabularnewline
		\noalign{\vskip0.2cm}
		{\footnotesize{}4.439} & {\footnotesize{}2.22 } & {\footnotesize{}5.88} &  & {\footnotesize{}1.10} & {\footnotesize{}2.37}\tabularnewline
		\noalign{\vskip0.2cm}
		{\footnotesize{}5.240} & {\footnotesize{}0.83 } & {\footnotesize{}0.57} &  & {\footnotesize{}1.28} & {\footnotesize{}1.18}\tabularnewline
		\noalign{\vskip0.2cm}
		{\footnotesize{}5.270} & {\footnotesize{}1.93 } & {\footnotesize{}1.39} &  & {\footnotesize{}3.27} & {\footnotesize{}3.25}\tabularnewline
		\noalign{\vskip0.2cm}
		{\footnotesize{}6.129} & {\footnotesize{}0.99 } & {\footnotesize{}0.83} &  & {\footnotesize{}1.69} & {\footnotesize{}0.84}\tabularnewline
		\hline 
		\noalign{\vskip0.2cm}
	\end{tabular}{\footnotesize{}\caption{Ratios of $\gamma$-ray production cross sections calculated by TALYS
		or derived from the Murphy et al. compilation \cite{R.J.MurphyB.KozlovskyJ.Kiener2009}
		to experimental data for two proton beam energies, E$_{p}$ = 66 and
		150 MeV.\label{tab:8}}
}
\end{table}
{\footnotesize \par}

\section*{Figures}

\begin{flushleft}
	\begin{figure}[H]
		\raggedright{}\includegraphics[scale=0.75]{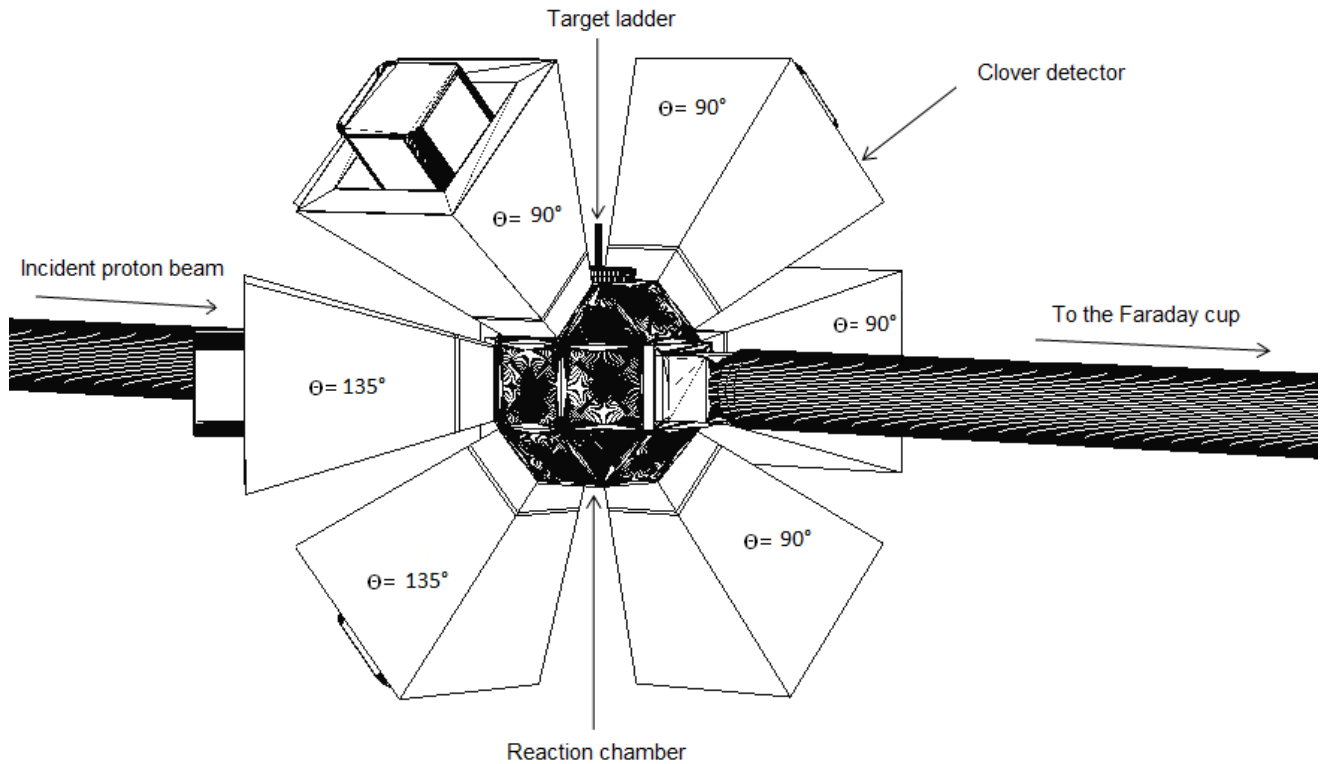}\caption{Schematic diagram of the experimental set up, showing the AFRODITE
			reaction chamber traversed by the incident proton beam within the
			beam pipe and the associated $\gamma$-ray clover detection array
			in the configuration used. \label{fig:1}}
	\end{figure}
	
	\par\end{flushleft}

\begin{flushleft}
	\begin{figure}[H]
		\raggedright{}\includegraphics[scale=0.95]{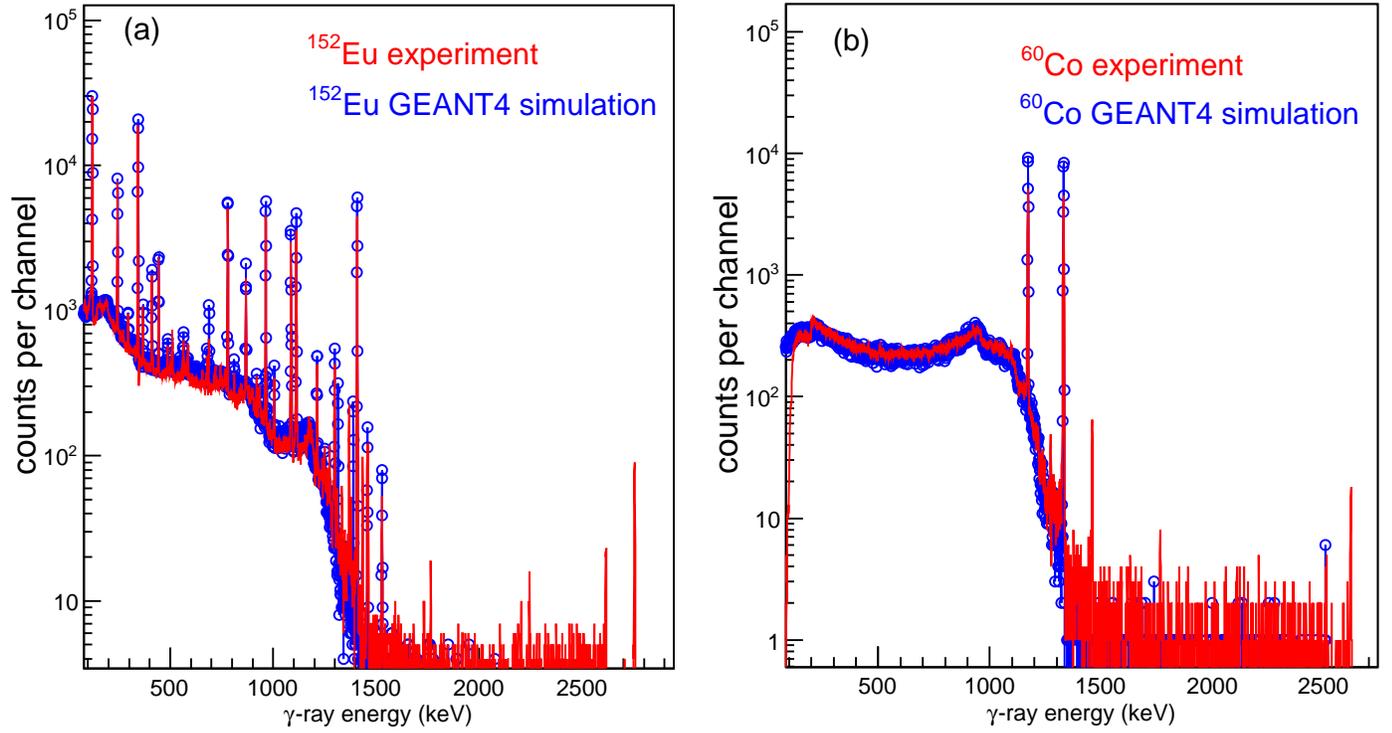}\caption{$\gamma$-ray energy spectra from radioactive sources placed at the
			target position, recorded by a HP-Ge detector located at $\theta_{lab}$
			= 95$\textdegree$: (a) $^{152}$Eu source and (b) $^{60}$Co
			source. The spectra resulting from a GEANT4 simulation with a detailed
			model of the experimental configuration and all the disintegration $\gamma$-ray
			lines from the $^{152}$Eu and $^{60}$Co radioactive
			sources are also plotted. The sum peaks from coincidence-cascades
			between full-absorption energy photons have negligible intensities
			relative to these lines both in the experimental and in the GEANT4-simulated
			energy spectra.\label{fig:2}}
	\end{figure}
	
	\par\end{flushleft}

\begin{flushleft}
	\begin{figure}[H]
		\raggedright{}\includegraphics[scale=0.9]{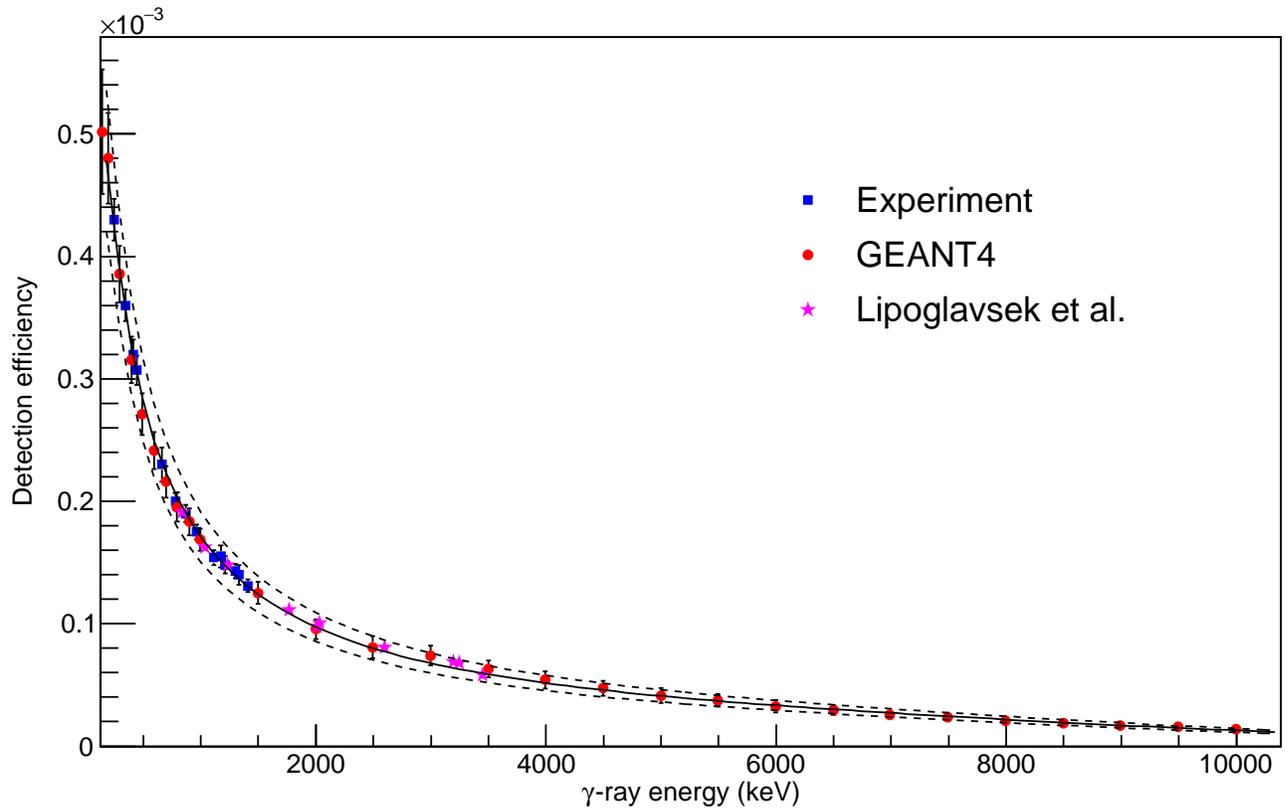}\caption{Experimental and GEANT4-simulated absolute detection efficiencies
			versus the $\gamma$-ray enegy from a single HP-Ge crystal placed at $\theta_{lab}$
			=85$^{{^\circ}}$ in the 2015 experiment. The fitted curve to the
			data is represented with a $\pm$ 12\% uncertainty shown by dotted
			lines. For the normalization of the simulated values to the experimental
			data, see the text in subsection \ref{sub:sub-section-A}. \label{fig:3}}
	\end{figure}
	
	\par\end{flushleft}

\begin{flushleft}
	\begin{figure}[H]
		\begin{raggedright}
			\includegraphics[scale=0.95]{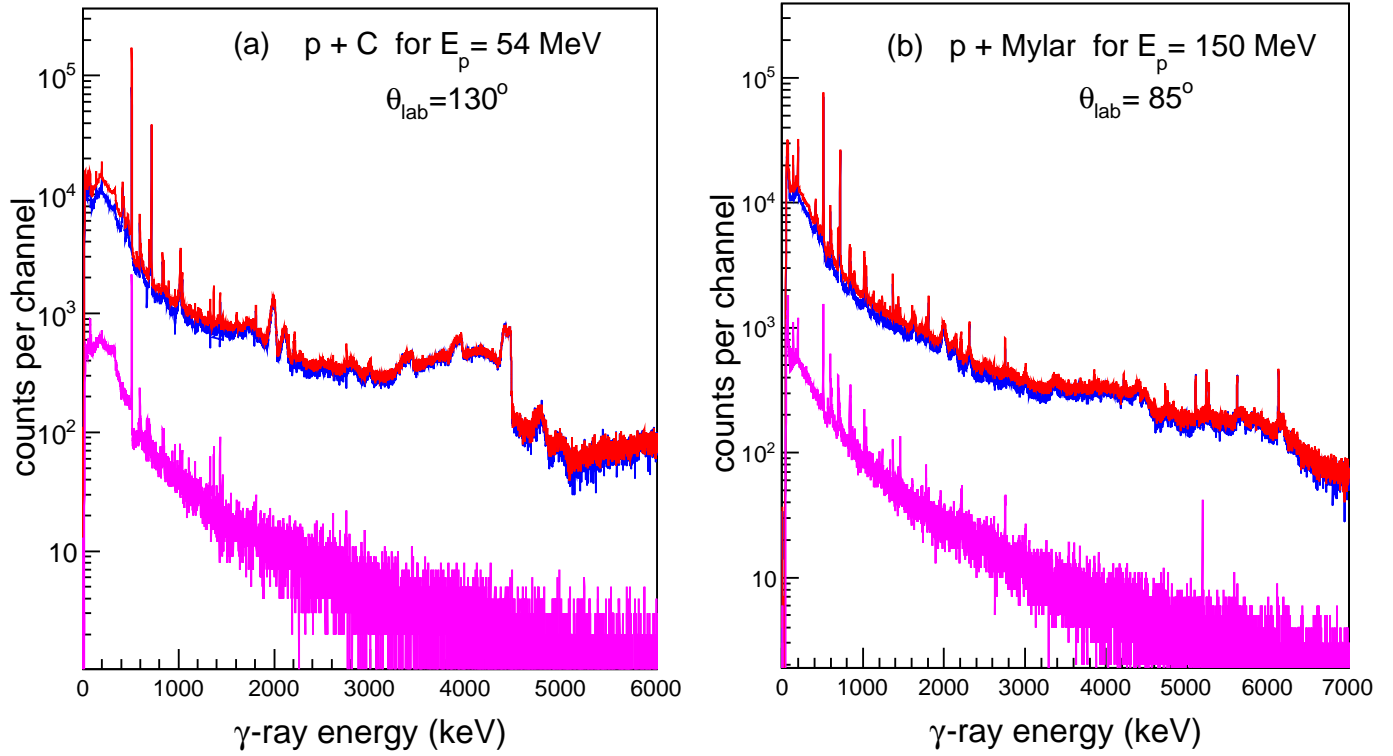}
			\par\end{raggedright}
		
		\raggedright{}\caption{$\gamma$-ray energy spectra recorded by two HP-Ge detectors in proton
			irradiations: (a) of the $^{nat}$C target for $\theta_{lab}$ = 130${^\circ}$,
			E$_{p}$ = 54 MeV, and (b) of the Mylar target for $\theta_{lab}$
			= 85${^\circ}$, E$_{p}$ = 150 MeV. Each figure comprises three spectra
			for the corresponding target: - in red color: raw experimental spectrum
			with target exposed to the proton beam - in purple: background spectrum
			with beam but without target in place, normalized to the same accumulated
			charge, - in blue: obtained net spectrum after background subtraction.\label{fig:4}}
	\end{figure}
	
	\par\end{flushleft}

\begin{flushleft}
	\begin{figure}[H]
		\raggedright{}\includegraphics[scale=0.95]{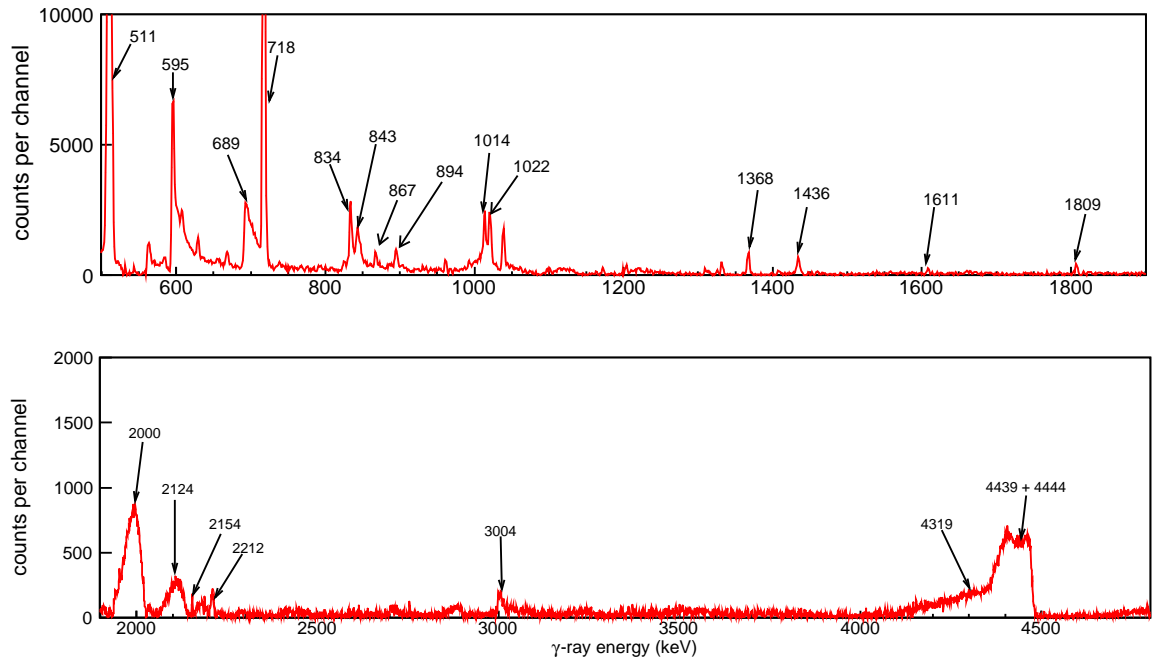}\caption{$\gamma$-ray energy spectra spectrum from an HP-Ge detector located
			at $\theta_{lab}$ = 130${^\circ}$ in the irradiation of the $^{nat}$C
			target with a proton beam of 54 MeV after subtraction of the Compton
			background and the components for the escape lines. The $\gamma$-ray
			lines of interest listed in TABLE \ref{eq:2} and the background lines
			from TABLE \ref{tab:4} are indicated by their corresponding energies
			in keV.\label{fig:5}}
	\end{figure}
	
	\par\end{flushleft}

\begin{flushleft}
	\begin{figure}[H]
		\raggedright{}\includegraphics{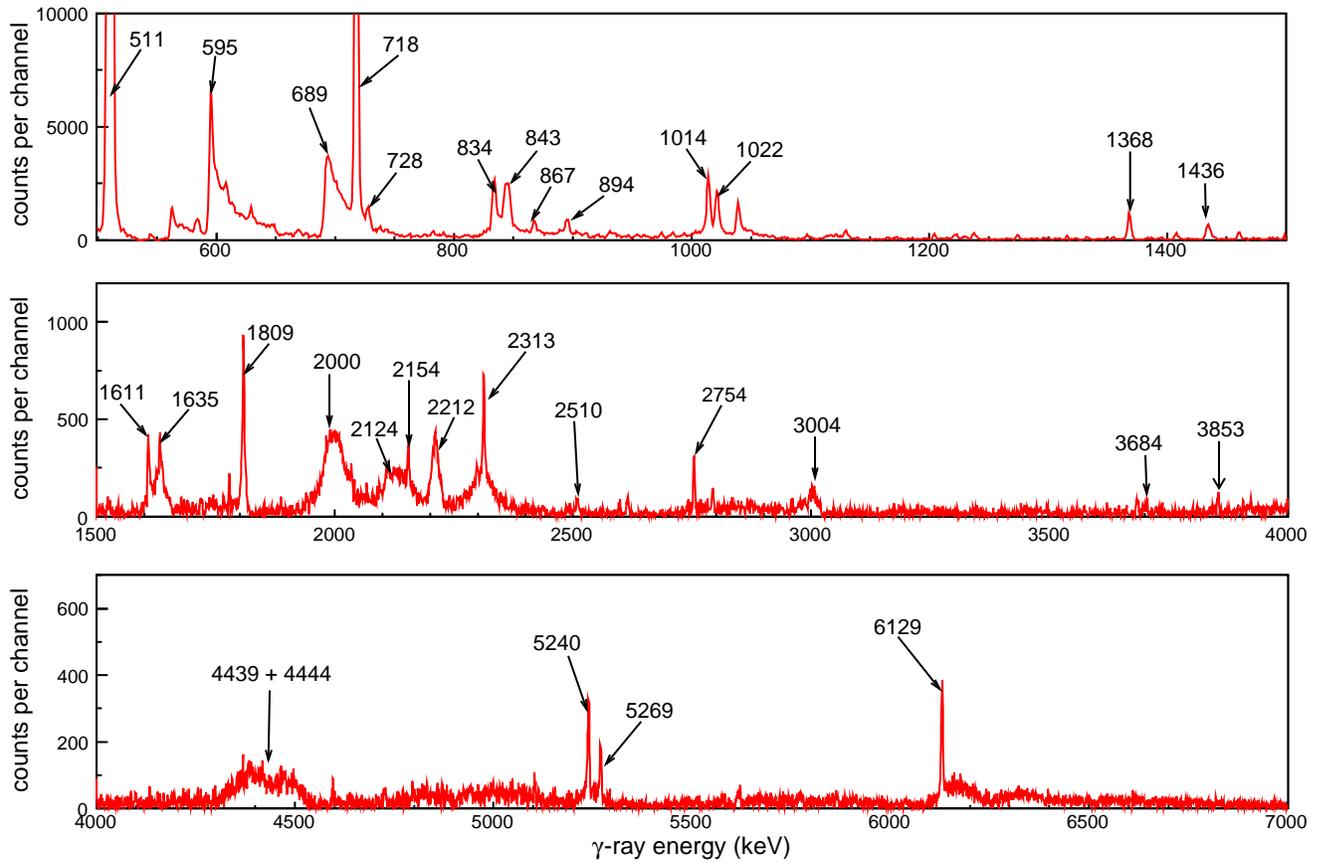}\caption{Similar to Fig. \ref{fig:5} for a HP-Ge detector located at $\theta_{lab}$
			=85${^\circ}$ but in irradiation of the Mylar target with a proton
			beam of 150 MeV. The $\gamma$-ray lines of interest listed in TABLES
			\ref{tab:2} and \ref{tab:3} and the background lines from TABLE
			\ref{tab:4} are indicated by their corresponding energies in keV.\label{fig:6}}
	\end{figure}
	
	\par\end{flushleft}

\begin{flushleft}
	\begin{figure}[H]
		\raggedright{}\includegraphics[scale=0.9]{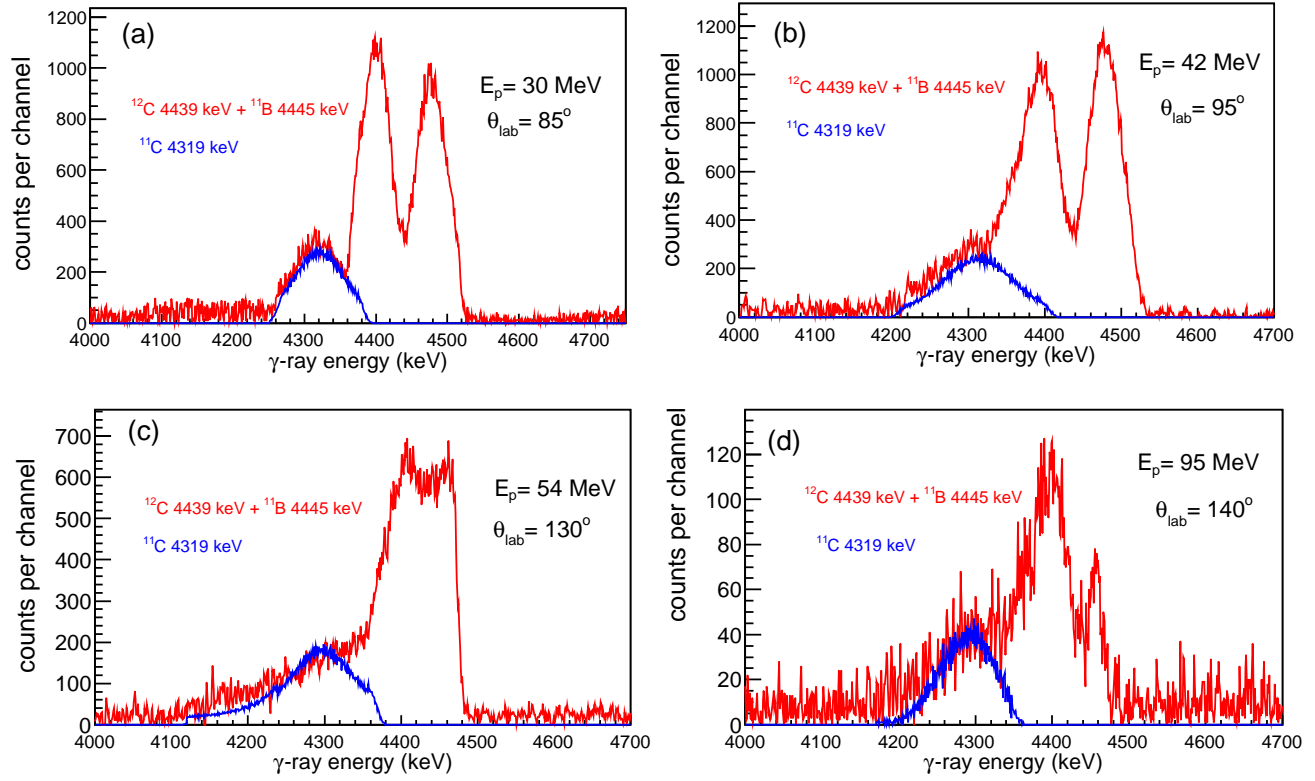}\caption{$\gamma$-ray profiles (counts per channel versus the proton beam
			energy E$_{p}$ and the observation angle $\theta_{lab}$) for the
			line complex observed at E$_{\gamma}$ = 4.44 MeV in proton irradiation
			of the $^{nat}$C target. The experimental profiles are depicted in
			red color, while the calculated profiles for the line of $^{11}$C
			at E$_{\gamma}$ = 4.319 MeV produced in inelastic proton scattering
			are represented in blue (see text).\label{fig:7}}
	\end{figure}
	
	\par\end{flushleft}

\begin{flushleft}
	\begin{figure}[H]
		\raggedright{}\includegraphics[scale=0.95]{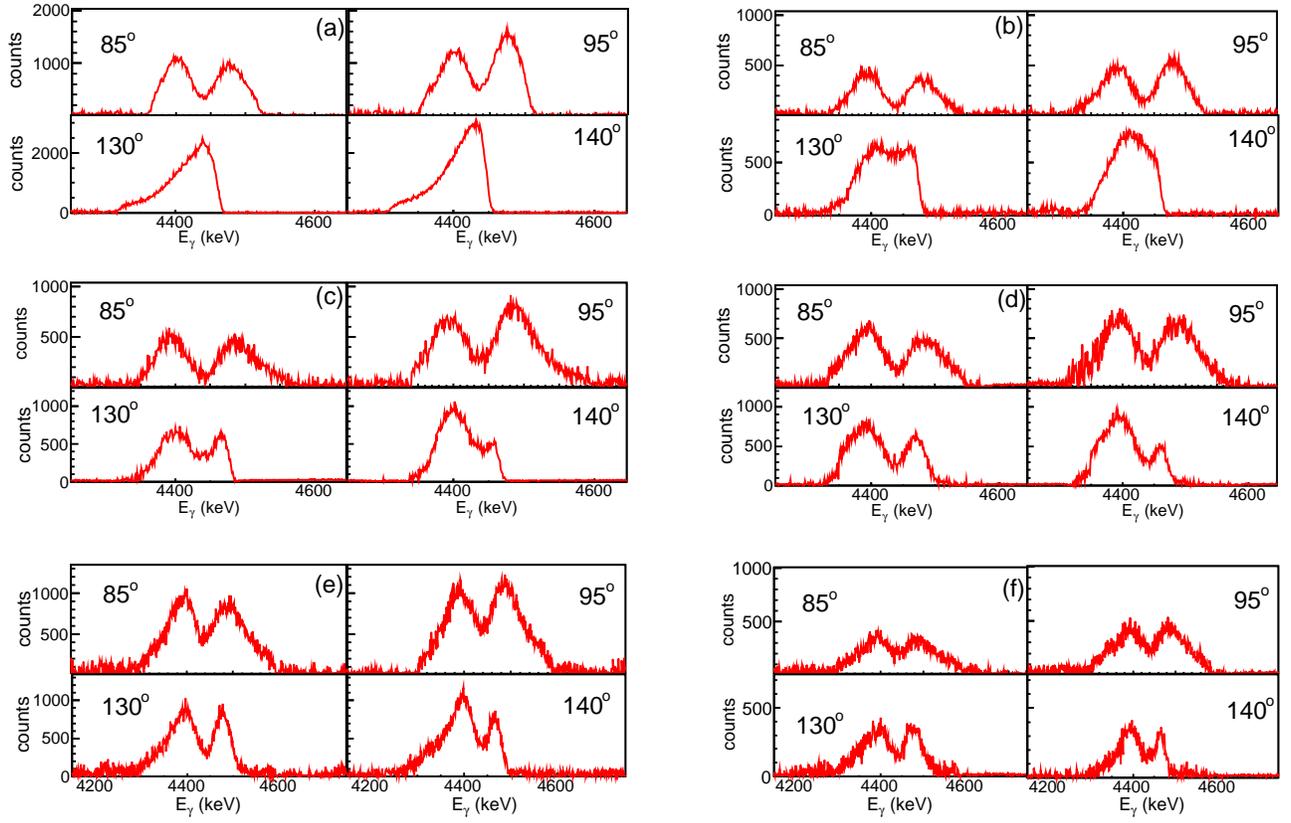}\caption{Measured profiles versus the proton beam energy (E$_{p}$) and the
			observation angle ($\theta_{lab}$) for the line complex at E$_{\gamma}$
			= 4.44 MeV after subtraction of the component for the 4.319 MeV line
			of $^{11}$C. The six panels (a), (b), (c), (d), (e) and (f) correspond
			to the proton beam energies E$_{p}$ = 30, 54, 80, 110, 150 and 175
			MeV, respectively.\label{fig:8}}
	\end{figure}
	
	\par\end{flushleft}

\begin{flushleft}
	\begin{figure}[H]
		\includegraphics[scale=0.95]{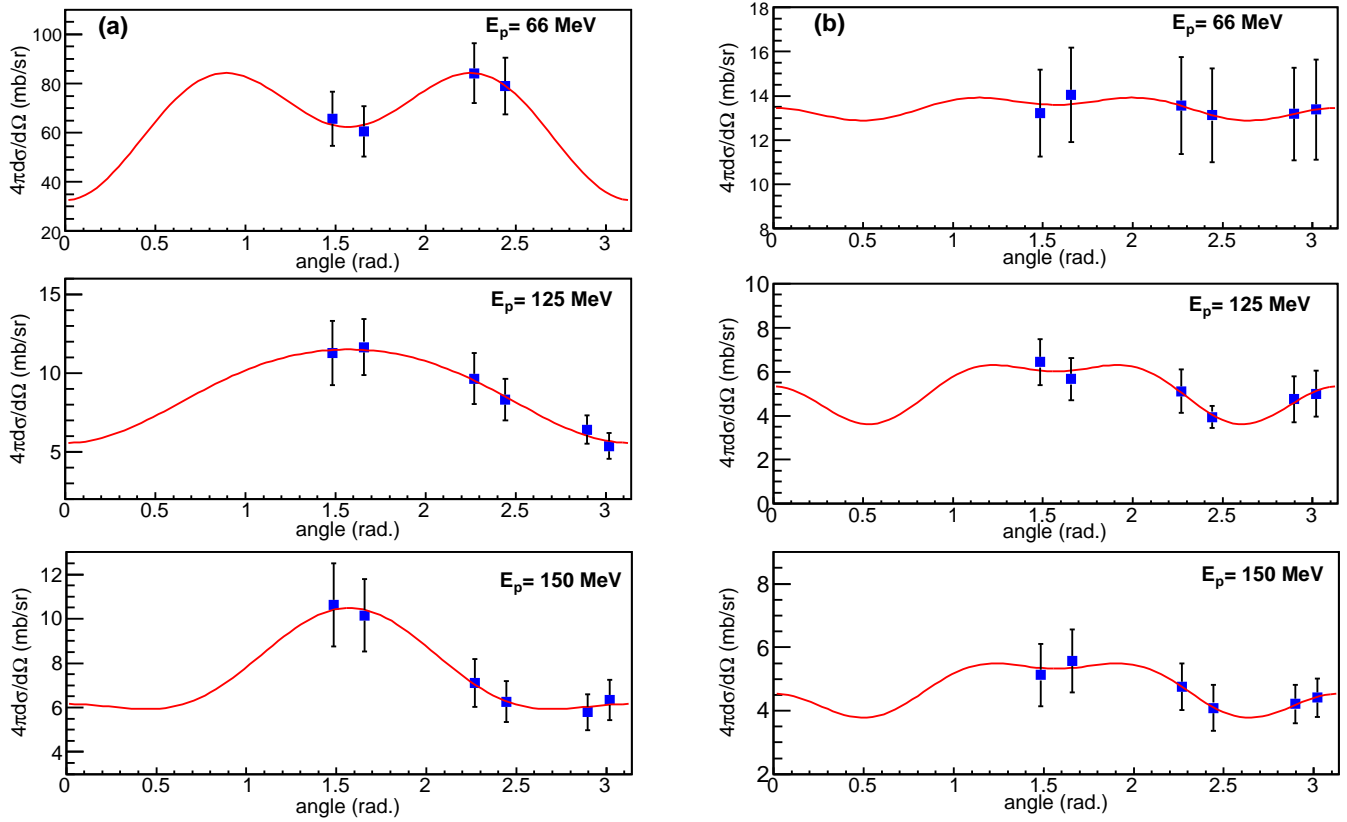}\caption{Examples of measured $\gamma$-ray differential cross section angular
			distributions (blue symbols) and Legendre polynomial expansion fits
			(red curves): (a) data for the line of $^{12}$C at E$_{\gamma}$
			= 4.44 MeV, (b) data for the line of $^{16}$O at E$_{\gamma}$ =
			6.129 MeV.\label{fig:9}}
	\end{figure}
	
	\par\end{flushleft}

\begin{flushleft}
	\begin{figure}[H]
		\raggedright{}\includegraphics[scale=0.95]{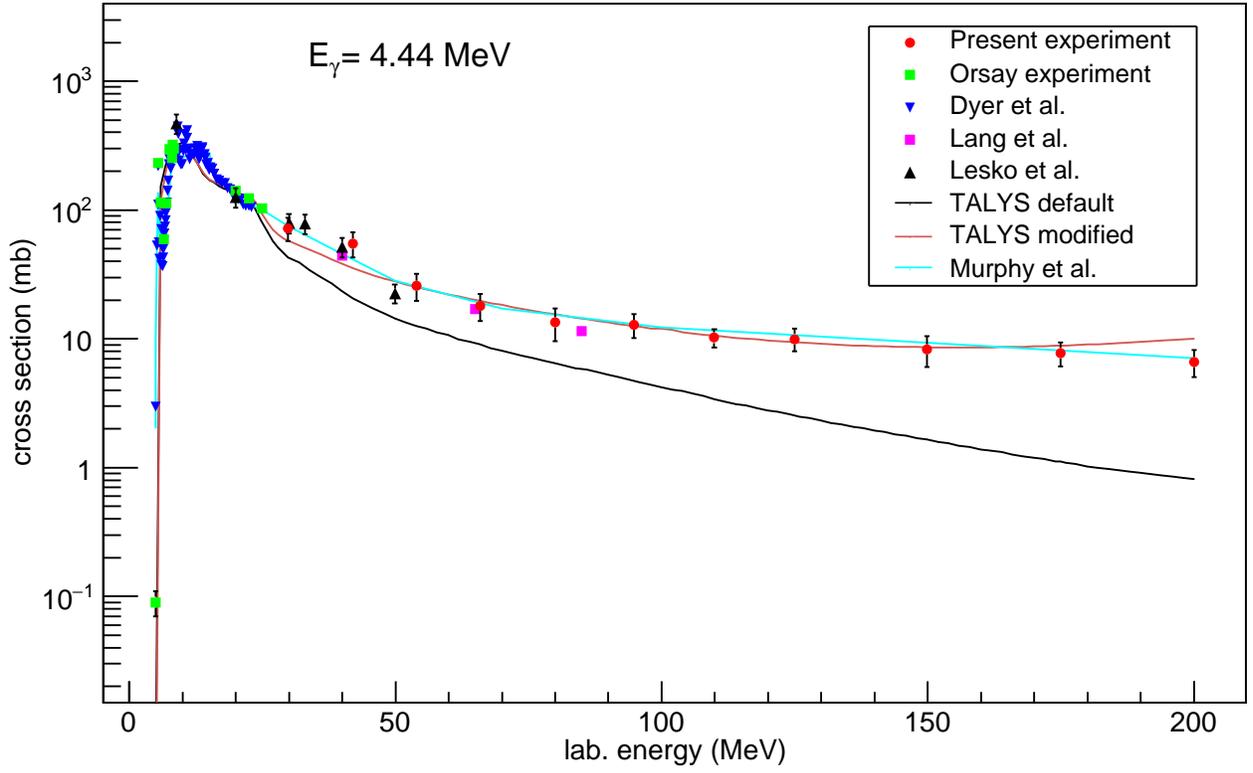}\caption{Integral cross section excitation functions for the 4.44-MeV $\gamma$-ray
			line produced in inelastic proton scattering off the $^{nat}$C target.
			The experimental data are shown by colored symbols: in red (this work),
			blue (Dyer et al.), green (Orsay group \cite{H.Benhabiles-Mezhoud2011,Belhout2007,Belhout2009,Kiener2008,Kiener1998}),
			magenta (Lang et al. \cite{Lang1987}), black (Lesko et al. \cite{Lesko1988}).
			The blue curve corresponds to the predictions of the Murphy et al.
			semi-empirical compilation (Ref. \cite{R.J.MurphyB.KozlovskyJ.Kiener2009}).
			The black curve represents the results derived by TALYS calculation
			with default input parameters, while the red curve refers to TALYS
			calculations with our modified OMP and $\beta_{\lambda}$ deformation
			parameters.\label{fig:10}}
	\end{figure}
	
	\par\end{flushleft}

\begin{flushleft}
	\begin{figure}[H]
		\begin{raggedright}
			\includegraphics[scale=0.95]{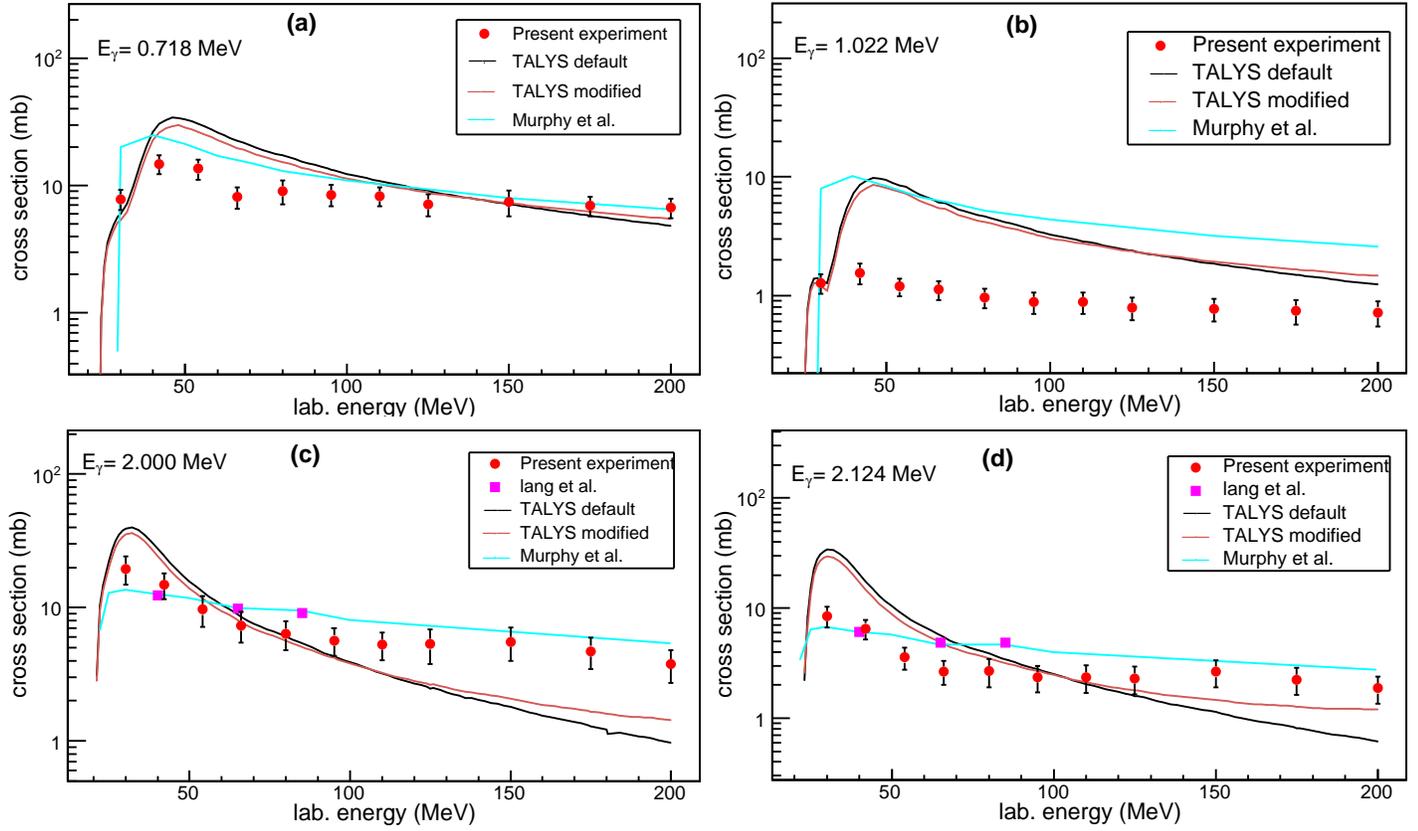}
			\par\end{raggedright}
		
		\caption{Excitation functions for four other $\gamma$-ray lines emitted in
			proton interactions with the $^{nat}$C target, using the same symbols
			as in Fig. \ref{fig:10}. Our experimental results for the lines of
			$^{11}$C and $^{11}$B at E$_{\gamma}$ = 2.000 and 2.124 MeV, respectively,
			are compared to Lang et al. \cite{Lang1987} data. \label{fig:11}}
	\end{figure}
	
	\par\end{flushleft}

\begin{flushleft}
	\begin{figure}[H]
		\raggedright{}\includegraphics[scale=0.95]{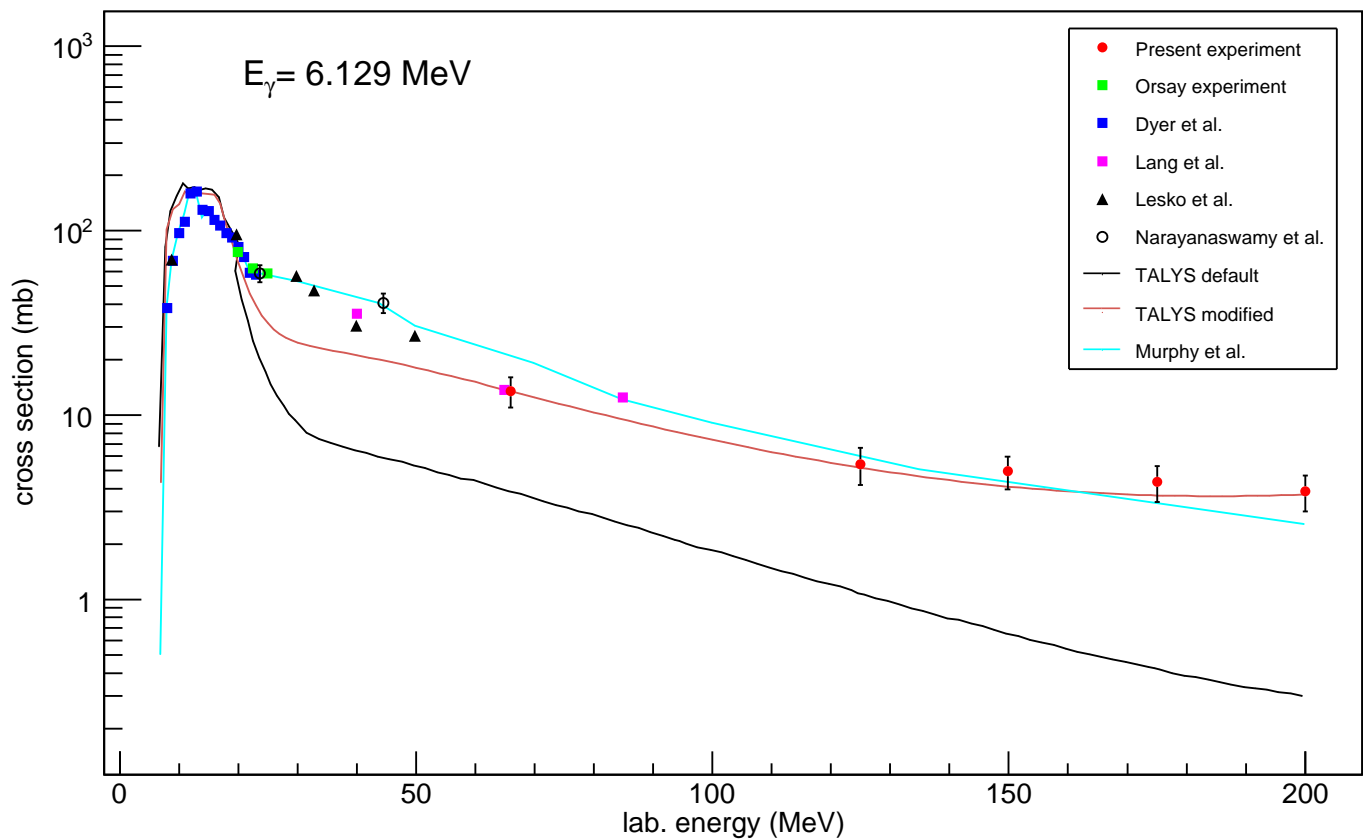}\caption{Same as in Fig. \ref{fig:10}, but for the 6.129-MeV main line of $^{16}$O
			produced in proton irradiation of the Mylar target. In addition, the
			black circles represent the Narayanaswamy et al. \cite{Narayanaswamy81}
			data.\label{fig:12}}
	\end{figure}
	
	\par\end{flushleft}

\begin{flushleft}
	\begin{figure}[H]
		\raggedright{}\includegraphics[scale=0.95]{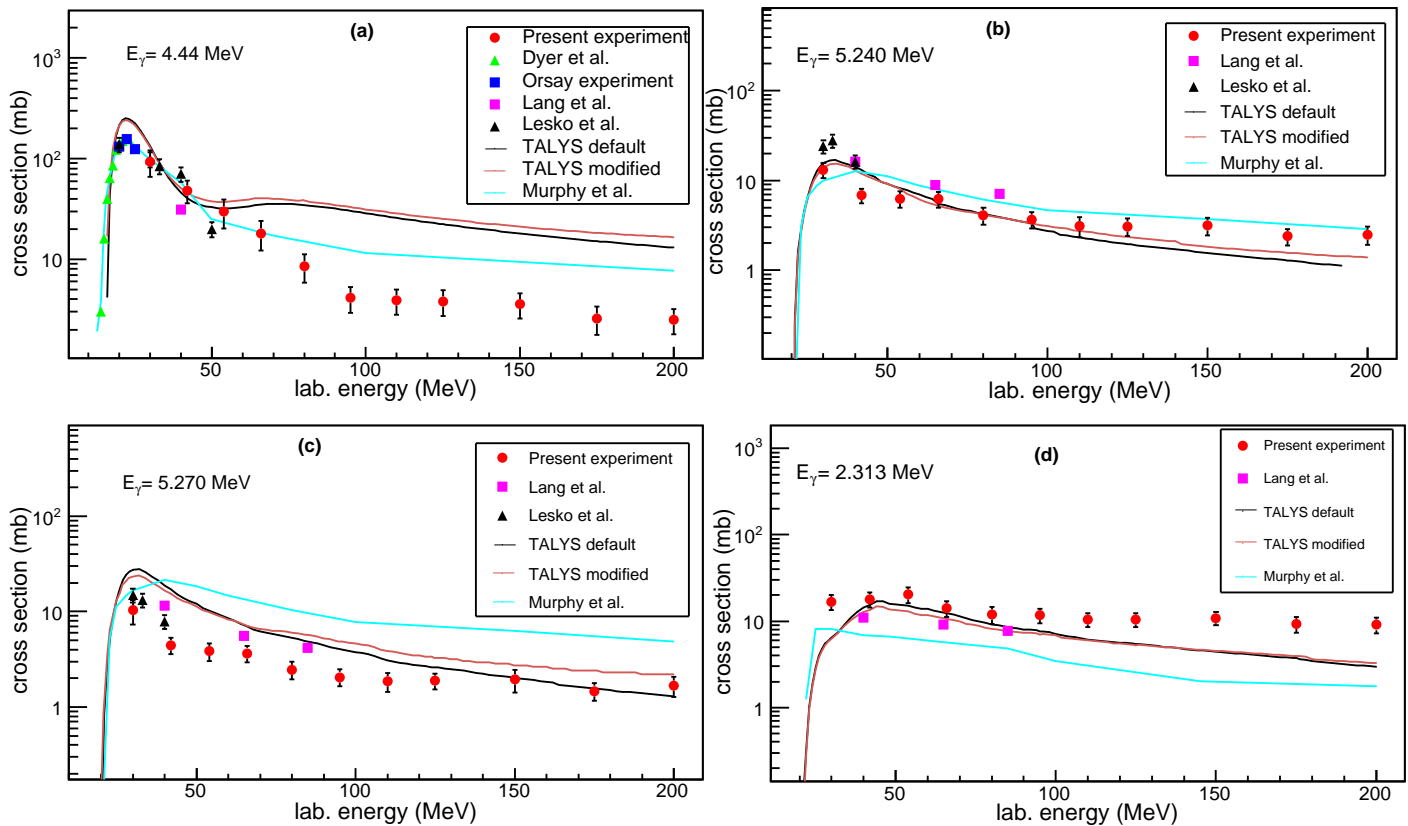}\caption{Same as in Fig. \ref{fig:11} but for four other $\gamma$-ray lines
			emitted in proton interactions with the Mylar target, using the same
			symbols as in Fig. \ref{fig:10}. Our experimental results are compared
			to existing previous data.\label{fig:13}}
	\end{figure}
	
	\par\end{flushleft}
\end{document}